\documentstyle[12pt]{article}
\global\arraycolsep=2pt
\newcommand{\k}{\vec{k}_{\perp}^2}
\newcommand{\be}{\begin{eqnarray}}
\newcommand{\ee}{\end{eqnarray}}
\newcommand{\la}{\langle}
\newcommand{\ra}{\rangle}

\newcommand{\Dm}{\vec{iD}_{\mu} }
\begin{document}

\begin{titlepage}
\begin{flushright}
SMU-HEP-94-9\\
hep-ph/9503476 \\
March 1995
\end{flushright}

\vspace{0.3cm}
\begin{center}
\Large\bf Transverse momentum distribution in hadrons.\\
  What can we learn from QCD ?
 \end{center}

\vspace {0.3cm}

 \begin{center} {\bf
Boris Chibisov\footnote{
e-mail addresses:boris@mail.physics.smu.edu}
and   Ariel R. Zhitnitsky\footnote{
On leave of absence from Budker Institute of Nuclear Physics,
Novosibirsk, 630090, Russia.
e-mail addresses: arz@mail.physics.smu.edu}}
 \end{center}

\begin{center}
{\it Physics Department, SMU , Dallas, Texas, 75275-0175}

\end{center}
\begin{abstract}

We discuss some  QCD constraints   on light-cone $\pi$ meson  wave
function   $\psi(\k, x)$ .
 The analysis is based on such general methods as    dispersion relations,
duality and  PCAC. We calculate the asymptotical behavior of the
wave function ($wf$)
at the end-point region ($x\rightarrow 1$ and  $\k\rightarrow\infty$)
by analysing  the corresponding  large  $n-$th  moments  in transverse
$\la\vec{k}_{\perp}^{2n}\ra \sim n!$ and   longitudinal
$\la (2x-1)^n\ra \sim 1/n^2$ directions. This information  fixes  the
asymptotic behavior of $wf$ at large $\k$ ( which
 is turned    out to be Gaussian commonly used in the phenomenological
analyses).

We discuss one particular application of the obtained results.
We calculate the nonleading ``soft"
contribution to the pion form factor at intermediate momentum
transfer.  We argue, that due to the
specific properties of   $\psi(\k, x)$, the corresponding contribution
can  temporarily {\bf simulate} the leading twist behavior
in the extent region of $ Q^2:~~3 GeV^2\leq Q^2\leq 40 GeV^2 $, where
$Q^2 F(Q^2)\sim const.$  Such a mechanism, if it is correct, would be
 an explanation of the phenomenological success of the dimensional
counting rules   at available, very modest energies
for many different processes. We discuss some inclusive amplitudes
( like Drell Yan and Deep Inelastic )
where intrinsic $\pi$ meson structure might be essential. The relation
to the valence  quark model   is also discussed.

\vskip 0.3cm
 \end{abstract}
\end{titlepage}

{\bf 1. Introduction}
 \vspace {0.3cm}

The main goal of the present paper is
the  study of
 the hadronic  wave functions   with a minimal number
of constituents
 within QCD. To be more specific, we are mainly interested in the $\k -$
behavior of the light cone $wf$ in the transverse direction.

The   motivation for this interest is  the following.
 As is known, at asymptotically high energies
  the  parametrically leading contributions
to hard exclusive processes can be expressed in terms of the so-called
distribution amplitudes   $\phi(x)$,
which itself can be expressed as an integral
$\int\psi (\k,x)d^2k$ with nonperturbative
wave function   $\psi(\k , x)$ ,
see review \cite{Brod1}.
 Distribution amplitudes  $\phi(x) $ depend only on longitudinal variables
$x_i$ and not on transverse $\k$ ones. The same is true
for inclusive reactions where structure functions depend on $x$,
but not on $\k$\footnote{ The formal reason for
that can be seen from the following arguments. At large energies the quark
and antiquark are produced at small distances $z\sim 1/Q\rightarrow 0$, where
$Q$ is typical
large
momentum transfer. Thus one can neglect the $z^2$
dependence everywhere and one should concentrate
on the one variable $zQ\sim 1$ which is order of one.
One can convince oneself that the standard Bjorken
variable $x$ is nothing but Fourier conjugated to $zQ$.}.
 Thus, any dependence on $\k$ gives
some power corrections to the leading terms.
Naively one may  expect that these corrections should be small
enough already in the   few $GeV^2$ region. However,
as we argue later, this expectation does  not seem work well
in intermediate region. Thus, one can say that we study   the pre -asymptotic
behavior of the exclusive amplitudes.

We shall find that $\psi (\k,x)$ possesses  the quite unusual properties,
which lead to the broadening $wf$ in the transverse directions. In terms
of observable amplitudes it means that the characteristic scale is not
$\sim 1 GeV^2$ (as naively one could expect), but $10 GeV^2$. What is more
important, this scale is not universal,
 but varies from process to process.

The analysis of some inclusive reactions
shows the same result-- very often the experimental data can not be explained
within the standard  scale-invariant
description.   An explicit introduction of some hadronic (phenomenological)
dimensional parameters into the structure functions is often required.
 Due to the fact that the structure functions can be expressed
in terms of the same  wave functions   as
$\sim \int |\psi (\k,x)|^2d^2k$,  we believe
that the similar conclusion
(on importance of the pre-asymptotic
behavior in the  intermediate region) takes place for the
inclusive amplitudes as well.

As the simplest application of our $wf$ we   consider
the $\pi$ meson form factor.
Before to explain the qualitative results   obtained in this paper,
we would like  to review a few important steps
which have been taken in the    investigation
of the exclusive amplitudes.  We hope that this short
 historical introduction will help to formulate the problem we want
to address in the present analysis.

 In early seventies the famous {\it dimensional
counting rules} were proposed \cite{Matveev}.
 The predictions of these rules
agree well with the experimental data, such as the
 pion and nucleon form factors,
large angle elastic scattering
cross sections and so on. This agreement served as a stimulus for
further theoretical investigations. The modern approach
to exclusive processes was started
  in the late seventies
and  early eighties  \cite{Brod}. We refer to the review papers
\cite{Brod1},\cite{Cher}, \cite{Cher1}
  for details .

The main idea of the approach \cite{Brod}
 is the separation of the large and small
distance physics. At small distances we can use the standard
perturbative expansion due to the asymptotic freedom and
smallness of the coupling constant. All nontrivial,
large distance physics is hidden into the
nonperturbative   $wf$  in this approach.
It can not be found
by perturbative technique, but rather  should be extracted
from  elsewhere.   The most powerful analytical
nonperturbative method
for such problems is the
QCD sum rules \cite{Shif1}, \cite{Shif2}.

The first application of QCD sum rules to the analysis
of nonperturbative $wf$ was considered more than decade ago
\cite{Cher2}.   Since then this subject
  is a very controversial
issue   \cite{Mikh}-\cite{Braun1} and we are not going
to comment  these quite opposite  points
in the present note.

At the same time, the applicability of the approach
\cite{Brod}  at experimentally
accessible momentum transfers was questioned \cite{Ditt},\cite{Isgur}.
In these papers
 it was demonstrated, that the perturbative, asymptotically leading
contribution, is much smaller than the nonleading ("soft")
contributions.
Similar conclusion, supporting this result, came from
the different side, from  the QCD sum rules,
\cite{Rad},\cite{Smilga}, where the direct
calculation of the form factor has been presented
 at $Q^2\leq 3GeV^2$.
  This method,
 has been extended later for the
  larger $  Q^2\leq 10 GeV^2$ \cite{Rad1},\cite{Braun1}
with the same qualitative result:
the soft contribution is more important in this intermediate region
than the leading one.

Now we are ready to formulate the question,
which we want to address in the present paper.

$\bullet$  If the asymptotically leading contribution can not provide
the experimentally observable absolute
values, than {\it how   can one
explain the very good agreement between the
experimental data and  dimensional counting rules}
\footnote{ These rules
unambiguously predict the dependence of amplitudes on dimensional
parameters. In particular, $Q^2 F_{\pi}(Q^2)\simeq constant.,~~
Q^4 F_{p}(Q^2)\simeq constant.,~~
s^7\frac{d\sigma}{dt}(\gamma p\rightarrow\pi^+n)=f(t/s)$
.. The experimental data are in a good agreement with
these predictions
in the large region of $s, Q^2 $ at very modest energy and momentum
transfer. },
which are supposed to be valid  only in the region where the
leading terms dominate?

It is clear, that the possible explanation can not be related to the
specific amplitude, but instead, it should be
connected, somehow, to the nonperturbative
wave functions  of the light hadrons ($\pi, \rho, p...$)
which enter  the formulae for exclusive processes.
The analysis of the $\pi$ meson
form factor, presented below supports this idea.

To anticipate the events we would like  to formulate here the result
of this analysis.
The very unusual properties
(which will be derived from QCD and not from quark model)
of the transverse momentum distribution
of the nonperturbative $\pi $ meson wave function lead to the
{\bf temporarily simulation} of the dimensional counting rules
by soft  mechanism for the $F_{\pi}(Q^2)$
at the extent range of intermediate momentum transfer:
$3 GeV^2\leq Q^2\leq 40 GeV^2 $. In this region   the soft contribution
 to $F_{\pi}(Q^2)Q^2 $ does not fall-off, as
naively one could expect, and we estimate it as
$F_{\pi}(Q^2)Q^2\simeq 0.3\div 0.4 GeV^2 $
 The leading twist contribution,
after Sudakov suppression, gives, according to
\cite{Kroll},\cite{Kiss},\cite{Braun1}
a little bit less (  $\leq 0.2 GeV^2 $).

$\bullet$ Therefore, our answer on the formulated question
is the  following.  The nonperturbative wave functions
possess (along with the standard small scale $\k\sim(330 MeV)^2$)
the new, larger scale ($\sim 1GeV^2$). Precisely
this new scale defines the regime where the asymptotical formulae
start to  work.

We believe that the same features of the $wf$ {\bf may affect}
the analysis of inclusive amplitudes as well
where some  pre-asymptotic effects might be essential.

The paper is organized as follows.
In the next section we define the nonperturbative $wf$ through its moments.
We   focus on the properties of the two particle leading twist $wf$
and its quark longitudinal and transverse distributions.
We recall our previous analysis
 regarding  the nonperturbative
$wf$ and formulate the main
constraints which have been obtained from QCD analysis.
In section 3 we model the $wf$ which satisfies these constraints.

Section 4 is
devoted to the calculation of the soft contribution to pion form factor
based on the model  $wf$s obtained in the
previous section. Let us stress from the very beginning:
we are not pretending to have made a reliable
calculation of the form factor here. We discuss some very general properties of
the
amplitudes
which are related to the specific features of the $\psi(\k, x)$.
We illustrate how these features change the behavior
of the form factor in intermediate region of $Q^2$.
 Section 5 is our conclusion and outlook.

\vspace {0.3cm}
{\bf 2.Constraints on the nonperturbative wave function $\psi(\k, x)$.}
 \vspace {0.3cm}

First of all let us review some essential definitions and
 results about nonperturbative $wf$.
We define the pion
axial wave function
in the following gauge-invariant way:
\be
\label{d}
if_{\pi}q_{\mu}\phi_A (zq,z^2)=
\la 0|\bar{d}(z)\gamma_{\mu}\gamma_5
e^{ig\int_{-z}^z A_{\mu}dz_{\mu}} u(-z)|\pi(q)\ra \\
\nonumber
=\sum_n \frac{i^n}{n!}\la 0|\bar{d}(0)\gamma_{\mu}\gamma_5
(iz_{\nu}\stackrel{\leftrightarrow}{D_{\nu}})^n u(0)|\pi(q)\ra ,
\ee
where
$\stackrel{\leftrightarrow}{D_{\nu}}\equiv
\stackrel{\rightarrow}{D_{\nu}}-\stackrel{\leftarrow}{D_{\nu}}$ and
$\Dm=i\vec{\partial_{\mu}}+gA_{\mu}^a\frac{\lambda^a}{2}$ is the
covariant derivative.
{}From its definition is clear that the set of different
$\pi$ meson matrix elements defines the nonperturbative wave
function.

The most important part (at  asymptotically high $q^2$) is
the one related to the longitudinal distribution.
In this case $z^2\simeq 0$  the $wf$
depends only on one $zq$- variable. The corresponding
Fourier transformed
wave function will be denoted as $\phi(\xi)$
and its
  $n-$th moment is given by the
following local matrix element:
\be
\label{1}
\la 0|\bar{d}\gamma_{\nu}
\gamma_5(i\stackrel{\leftrightarrow}{D_{\mu}}
z_{\mu})^{n} u|\pi(q)\ra=if_{\pi}q_{\nu}
(zq)^{n} \la \xi^{n}\ra=if_{\pi}q_{\nu}(zq)^{n}
\int^1_{-1}d\xi\xi^{n}\phi(\xi)
\ee
\be
\label{}
-q^2\rightarrow\infty,~~zq\sim 1~~
\xi=x_1-x_2,~~ x_1+x_2=1,~~z^2=0.   \nonumber
\ee
 Therefore, if we knew all matrix elements (\ref{1})
( which are well-defined ) we could restore the whole distribution amplitude
$\phi(\xi)$.
 The QCD sum rules approach allows one to find the
magnitudes only the few first moments \cite{Cher2}.
As is known, this information is  not enough to reconstruct
the $wf$; the parametric behavior at $\xi\rightarrow\pm 1$
is the crucial issue in this reconstruction.

To extract the corresponding information,
we use the following duality argument.
Instead of consideration of the  pion $wf$ itself, we study
  the following correlation function with pion quantum numbers:
\be
\label{2}
i  \int dx e^{iqx}\la 0|T J_{n}^{\|}(x),J_0(0) |0\ra=
(zq)^{n+2}I_{n}(q^2),~~
 J_{n}^{\|}=\bar{d}\gamma_{\nu}z_{\nu}
\gamma_5(i\stackrel{\leftrightarrow}{D_{\mu}}z_{\mu})^{n} u
 \ee
and calculate its asymptotic behavior at large $q^2$.
The result can be presented in the form of the dispersion integral,
whose spectral density is determined by the pure perturbative one-loop
diagram:
\be
\label{3}
\frac{1}{\pi}\int_0^{\infty} ds\frac{Im I_n^{pert}(s)}{s-q^2},~~
 Im I_{n}(s)^{pert}=\frac{3}{4\pi(n+1)(n+3)}.
 \ee
  We {\bf assume } that the $\pi$ meson gives a
nonzero contribution to the dispersion integral for arbitrary
$n$ and, in particular, for $n\rightarrow\infty$.
Formally, it can be written in the following way
 \be
\label{3a}
 \frac{1}{\pi}\int_0^{S_{\pi}^n} ds Im I(s)^{pert}_{n}=
\frac{1}{\pi}\int_0^{\infty} ds Im I(s)^{\pi}_{n},
 \ee
   Our assumption means
  that there are no special cancelations
and $\pi$ meson contribution to the
dispersion integral is not zero, i.e. $S_{\pi}^n(\|)\neq 0$,
where we specified the notation for the longitudinal distribution.
In this case at $q^2\rightarrow\infty$ our assumption (\ref{3a})
leads to the following relation:
\be
\label{4}
  f_{\pi}^2\la \xi^n\ra (n\rightarrow\infty)
\rightarrow\frac{3S_{\pi}^{\infty}(\|)}{4\pi^2n^2}
 \ee
It  unambiguously  implies the  following behavior at
the end-point region \cite{Cher}:
 \be
\label{5}
 \la\xi^n\ra=\int_{-1}^1d\xi \xi^n\phi(\xi)\sim 1/n^2,~~~~~
   \phi(\xi\rightarrow
\pm 1)\rightarrow (1-\xi^2).
\ee
Few comments are in order. We consider the nonperturbative
correlation function (\ref{2}). Thus the behavior (\ref{4})
should be fulfilled for any nonperturbative $wf$. The  perturbative
as well as nonperturbative corrections will change the duality
interval $S_{\pi}^n(\|) $
in the formula (\ref{4}) in comparision with perturbative one-loop
calculation. However,  $1/n^2$- behavior remains unaffected.

 Thus, our first  constraint   looks as follows:

$\bullet 1~~~~~~~~~~~~~~~~~~~~~~~
   \phi(\xi\rightarrow
\pm 1)\rightarrow (1-\xi^2)$.

We want to emphasize that
we did not use any numerical approximation
in this derivation. Therefore, the constraint ($\bullet 1$)
 has very general origin
 and it should be considered
as a direct consequence of QCD. Only dispersion relations, duality
and very plausible assumption formulated above have been used in
the derivation ($\bullet 1$).

 Now we want to repeat these arguments for the analysis of the
transverse distribution. To do so, let us define
the mean values of the transverse quark distribution
by the following matrix elements:
\be
\label{6}
\la 0|\bar{d}\gamma_{\nu}
\gamma_5 (\stackrel{\rightarrow}{iD_{\mu}}
 t_{\mu})^{2n} u|\pi(q)\ra=if_{\pi}q_{\nu}
 (-t^2)^n\frac{(2n-1)!!}{(2n)!!}\la \vec{k}_{\perp}^{2n} \ra.
\ee
where $\stackrel{\rightarrow}{D_{\nu}}$ is the
covariant derivative,
acting on the one quark  and
 transverse vector $t_{\mu}=(0,\vec{t},0)$ is perpendicular
 to the hadron momentum $q_{\mu}=(q_0,0_{\perp},q_z)$.
The factor $\frac{(2n-1)!!}{(2n)!!}$ is introduced   to
(\ref{6}) to take into account
  the integration over $\phi$ angle in the transverse plane:
$\int d\phi (\cos\phi)^{2n}/ \int d\phi= {(2n-1)!!}/{(2n)!!}$.

We interpret the $\la \vec{k}_{\perp}^{2} \ra$ in this equation
as  a mean value of the quark perpendicular momentum. Of course it
is different from the naive, gauge dependent  definition like
$\la 0|\bar{d}\gamma_{\nu}
\gamma_5 \partial_{\perp}^2 u|\pi(q)\ra $,
because the physical transverse gluon is participant
of this definition.
However, the expression (\ref{6}) is the only possible
way to define the  $\k$
in the gauge theory like QCD.
 We believe that such definition is the useful generalization
of the transverse momentum conception for the interactive quark
system.

  To find the behavior $\la \vec{k}_{\perp}^{2n} \ra$ at large $n$
  we can repeat our previous
duality arguments with the following result
\footnote{ Here
and in what follows we ignore any mild (nonfactorial) $n$-dependence.}:
\be
\label{8}
f_{\pi}^2\la \vec{k}_{\perp}^{2n} \ra \frac{(2n-1)!!}{(2n)!!}\sim
n!\Rightarrow
 f_{\pi}^2\la \vec{k}_{\perp}^{2n} \ra \sim n!.
\ee
 This behavior has been obtained in ref.\cite{Zhit2}
by analysing the perturbative series  of the specific
correlation function at large order. The dispersion relations
and duality arguments translate this information into the
formula (\ref{8}). Let us repeat again:
any nonperturbative wave function should respect  eq.(\ref{8})
 in spite of the fact that
apparently we calculate only  the perturbative part
( see
comment after formula (\ref{5})).

 The nice feature of   (\ref{8})
is its finiteness  for arbitrary $n$. It means
that the higher  moments
$$
\la \vec{k}_{\perp}^{2n} \ra =
\int d\k d\xi
\vec{k}_{\perp}^{2n}\psi(\k, \xi ) $$
{\bf do exist}.
In this formula we introduced the nonperturbative $wf$
$\psi(\k, \xi )$, normalized to one.
Its moments are determined by the local
matrix elements (\ref{6}). The relations to Brodsky and Lepage
notations $\Psi_{BL}(x_1,\vec{k}_{\perp})$
\cite{Brod1} and to longitudinal distribution
amplitude $\phi(\xi)$ introduced earlier, look as follow:
\be
\label{9}
\Psi_{BL}(x_1,\vec{k}_{\perp})=\frac{f_{\pi}16\pi^2
}{\sqrt{6}}\psi(\xi,\vec{k}_{\perp}),~
\int d\k \psi(\k, \xi )
= \phi(\xi),~\int_{-1}^1 d\xi\phi(\xi)=1
\ee
where $f_{\pi}=133 MeV$.
The  existence of the arbitrary high moments
$\la \vec{k}_{\perp}^{2n} \ra$ means that the nonperturbative
$wf$, defined above, falls off at large transverse momentum $\k$
faster than any power function.
   The relation (\ref{8})
fixes the asymptotic behavior of $wf$ at large $\k$.
Thus, we arrive to the following constraint:

$\bullet 2~~~~~~~~~~~~~~~~
   \la \vec{k}_{\perp}^{2n} \ra =
\int d\k d \xi
\vec{k}_{\perp}^{2n}\psi(\k, \xi )\sim n!
 ~~~~~n\rightarrow\infty  .$

We can repeat our duality arguments again for an arbitrary number of
transverse derivatives and large ($n\rightarrow\infty$) number
of longitudinal derivatives with the following result \cite{Zhit1}:

$\bullet 3~~~~~~~~~~~~~~~~~~
  \int d\k
\vec{k}_{\perp}^{2k}\psi(\k, \xi\rightarrow\pm 1 )\sim  (1-\xi^2)^{k+1}.$

For the $k=0$ we reproduce our previous formula
for the $\phi$   function: $\phi (\xi\rightarrow\pm 1)=
\int d\k \psi(\k, \xi\rightarrow\pm 1)\sim (1-\xi^2) $.
The constraint ($\bullet 3$)
 is extremely important and implies that the $\k$
dependence of the   $\psi(\k, \xi )$ comes
{\bf exclusively in the combination}
$ \k/(1-\xi^2)$ at $\xi\rightarrow\pm 1$.  The byproduct of this
constraint can be formulated as follows. The standard assumption
on factorizability of the $\psi(\k, \xi ) =\psi(\k )\phi(\xi)$
{\bf does contradict} to the very general properties of the theory.
 Thus, the asymptotic behavior of the $wf$   turns out to be Gaussian one
with the very specific argument:
\be
\label{!}
 \psi(\k\rightarrow\infty, \xi)\sim \exp(-\frac{\k}{(1-\xi^2)} ),
 \ee
 We would like to pause here in order to make the following conjecture.
The Gaussian $wf$ (reconstructed above from the  QCD analysis)
 not accidentally coincides with the harmonic oscillator
$wf$ from constituent quark model.
 To make this conjecture more clear,    let us recall
 few results  from the constituent quark model.

It is well known \cite {Isgur1} that the equal- time
 wave functions
\be
\label{qm}
\psi_{CM}(\vec{q}\,^2)\sim \exp (-\vec{q}\,^2)
\ee
of the harmonic oscillator in the rest frame
give a very reasonable description of
static meson properties.
Together with Brodsky-Huang-Lepage prescription \cite{Terentyev},
\cite{BHL}
connecting  the equal -time   and the
light-cone wave functions of two  constituents (with mass $m\sim  300 MeV$)
by identification
$$ \vec{q}^2\leftrightarrow\frac{\k+m^2}{4x(1-x)}-m^2,
{}~~\psi_{CM}(\vec{q}^2)\leftrightarrow\psi_{LC}
(\frac{\k+m^2}{4x(1-x)}-m^2),$$
one can reproduce the
Gaussian behavior (\ref{!}) found  from QCD.
It means, first of all, that our identification
of the moments (\ref{6}) defined in  QCD
with the ones defined in quark model,
is the reasonable conjecture
\footnote{The same method can be applied for the analysis
of the asymptotical behavior of the nucleon $wf$ which
in obvious notations takes the form:
$$\psi_{nucleon}(\vec{k}_{\perp i}^2
\rightarrow\infty, x_i)\sim \exp(-\sum\frac{ \vec{k}_{\perp i}^2}{x_i})$$.}.

However, there is a difference. In quark model
we do have a parameter which describes the mass
of constituent $m\simeq 300 MeV$. We have nothing like that
in QCD. This difference has very important consequences
 which will be discussed later.

We would like to put a few more constraints on the list.
But before to do so, we have to emphasize  the difference
between  constraints
($\bullet 1-\bullet 3$) discussed above
and  the ones which follow. The first three constraints
have  very general origin. No numerical approximations have been made
in the derivation of the corresponding formulae. The only what
 have been used are dispersion relations, duality and very plausible
 assumption formulated above. The main idea of the derivation of  all these
constraints
is one and the same: we calculate some correlation function in QCD
using the asymptotic freedom. The dispersion relations and duality
arguments transform these properties into the
constraints on  hadronic matrix elements.

The constraints we are going to discuss now have
absolutely different status. They are based on the QCD sum rules
with their inevitable numerical assumptions about  higher
excited states in QCD. Thus, they must be treated as
an approximate ones. The well-known constraint of such a kind is the second
moment of the distribution
amplitude in  the
longitudinal direction \cite{Cher2}:

$\bullet 4 ~~~~~~~~~~~~~~~~~~~~~~
\la\xi^2\ra \equiv \int d\xi \phi(\xi)\xi^2\simeq 0.4,$

 (The asymptotic $wf$ corresponds to  $\la\xi^2\ra\ = 0.2$ ).
Such a result was the reason to suggest   the ``two-hump" shape $wf$
\cite{Cher2}
which meets the above  requirement. The number
cited as  the constraint ($\bullet 4$)
has been seriously criticized  in refs. \cite{Mikh}-\cite{Rady}.
The point for criticism was exactly the
assumption about the role of the excited states in the sum rules.
We can not answer on this criticism within standard QCD sum rules approach.
Thus, in what follows we shall discuss both possibilities: the narrow
(asymptotic)  $wf$  and  the wider one (with larger $\la\xi^2\ra\ >0.2$).

The next ``numerical" constraint
is the second moment of the $wf$
in the transverse direction defined by equation
(\ref{6}) and calculated for the  first time in \cite{Cher} and
independently (with quite different technique) in \cite{Novik}.
Both results are in a full agreement to each other:

$\bullet 5 ~~
 \la \vec{k}_{\perp}^{2} \ra=\frac{5}{36}\frac{\la
\bar{q}ig\sigma_{\mu\nu}
 G_{\mu\nu}^a\frac{\lambda^a}{2} q \ra }{\la \bar{q}q\ra}
\simeq \frac{5 m_0^2}{36}\simeq 0.1GeV^2, ~~~m_0^2\simeq 0.8 GeV^2.$

Essentially, the constraint ($\bullet 5$) defines the general scale
of all nonperturbative phenomena for the  pion. It is not
accidentally coincides with $ 300 MeV$   which is the
typical magnitude in the hadronic physics.

All numerical values obtained within QCD sum rules approach
correspond to the normalization point $\sim 1GeV^2$.
To model the $wf$ we need to know their renormalization
properties.
 The  anomalous dimensions of the longitudinal operators are
well known, see e.g.\cite{Brod1}, \cite{Cher}. For our particular case
we can write it in the following way:
\be
\label{10}
(\la\xi^2\ra-\frac{1}{5})_{\mu_1}
= (\frac{\alpha_s(\mu_1)}{\alpha_s(\mu_2)})^{50/9b}\cdot
(\la\xi^2\ra-\frac{1}{5})_{\mu_2},~~~~b=\frac{11}{3}N_c-\frac{2}{3}n_f=9
 \ee

The anomalous dimensions of the higher twist operators
(which are related to transverse moments) are less familiar.
 It has been calculated for
  the operator   describing the mean value of the transverse momentum
$\k$ \cite{Braun}:
\be
\label{11}
\la\k\ra_{ \mu_1}
= (\frac{\alpha_s(\mu_1)}{\alpha_s(\mu_2)})^{\frac{32}{9b}}
\cdot\la\k\ra_{\mu_2}.
\ee
 To study  the fine properties of the
transverse distribution it is desired to know the next moment.
The problem can be reduced to the analysis of the
 mixed vacuum condensates of dimension seven \cite{Zhit1}:
 \be
\label{12}
  \la\vec{k}_{\perp}^{4}\ra =
 \frac{1}{8}\{   \frac{-3\la \bar{q}g^2\sigma_{\mu\nu}
 G_{\mu\nu} \sigma_{\lambda\sigma}
 G_{\lambda\sigma}q \ra}{4\la\bar{q}q\ra}
+ \frac{ 13\la \bar{q}g^2
 G_{\mu\nu}   G_{\mu\nu}q \ra}{ 9 \la\bar{q}q\ra}\},
\ee
We analyzed the magnitudes for  these vacuum condensates with
the following result: the standard factorization hypothesis does not work
in this case. The factor of nonfactorizability $K\simeq 3.0\div 3.5$
\cite{Zhit1}, \cite{Zhit3}.
The eq.(\ref{12}) defines the new numerical constraint on the transverse
distribution.
We prefer to express this constraint not in  terms of the absolute
values, but rather, in terms of the dimensionless parameter $R$
which is defined in the following way:

$\bullet6 ~~~~~~~~~ R\equiv
\frac{\la \vec{k}_{\perp}^{4}\ra  }{\la \vec{k}_{\perp}^{2}\ra^2}
\simeq 3K\cdot\frac{ \la g^2G_{\mu\nu}^aG_{\mu\nu}^a\ra}{m_0^4} \simeq  5\div
7,
{}~m_0^2\simeq 0.8GeV^2    ,$

 where we use the standard values for parameter $m_0^2$ and gluon
condensate \cite{Shif2}. We would like to emphasize
that the fluctuations of the transverse momentum are large enough. The
quantitative
 characteristic of these fluctuations is parameter
$R\gg1$. In terms of wave function this property means a very
unhomogeneous distribution in transverse direction.

As we already mentioned, to model $wf$ we need to know
the renormalization properties  of the higher twist
operators. Unfortunately, we do not know them (except for
the lowest one, (\ref{11})). However, one can argue,
that in the large $N_c$ limit, the main contribution to anomalous
 dimensions can be found from the formulae like (\ref{12}),
where matrix elements are expressed in terms of vacuum condensates.
In the same $N_c\rightarrow\infty$ limit the renormalization
properties of these condensates (but not their absolute values)
can be estimated (with the same accuracy)
 by applying the factorization procedure.
Thus,  one could expect that this prescription
gives a reasonable numerical accuracy ($\sim 1/N_c$) for
the renormalization properties of higher twist
operators.
However, the exact calculations are highly desired
and welcome \cite{Chibisov}.

We would like to check this prescription for the   operator
 with known   dimension (\ref{11}).
The anomalous dimensions for  the chiral condensate
and mixed condensate are known:
\be
\label{13}
\la \bar{q}q\ra_{\mu_1}=(\frac{\alpha_s(\mu_1)}
{\alpha_s(\mu_2)})^{\frac{-4 }{b}}\cdot\la \bar{q}q\ra_{\mu_2}
\ee
\be
  \la \bar{q}ig\sigma_{\mu\nu}
 G_{\mu\nu}^a\frac{\lambda^a}{2} q \ra_{\mu_1}=
(\frac{\alpha_s(\mu_1)}{\alpha_s(\mu_2)})^{\frac{-2 }{3b}}\cdot
\la \bar{q}ig\sigma_{\mu\nu}
 G_{\mu\nu}^a\frac{\lambda^a}{2} q \ra_{\mu_2}  \nonumber
 \ee
Thus, our relation ($ \bullet 5$)   gives the following
prescription  for the evolution formula under the renormalization
group transformation:
\be
\label{14}
\la \vec{k}_{\perp}^{2} \ra_{\mu_1}\sim\frac{\la
\bar{q}ig\sigma_{\mu\nu}
 G_{\mu\nu}^a\frac{\lambda^a}{2} q \ra }{\la \bar{q}q\ra}
\sim(\frac{\alpha_s(\mu_1)}{\alpha_s(\mu_2)})^{\frac{30}{9b}}
\cdot\la \vec{k}_{\perp}^{2} \ra_{\mu_1},
\ee
 instead of exact formula (\ref{11}). We consider this
  good numerical agreement as a justification for the analogous
estimation   for   different operators
with unknown anomalous dimensions. In particular,
we expect that the matrix element
$\la\vec{k}_{\perp}^{4}\ra_{\mu}$ (\ref{12}) is not changed
strongly under the renormalization group transformations
because after the factorization this operator reduces
to the gluon condensate which is renormalization invariant.

At the same time our dimensionless parameter $R$ is changed
strongly and we estimate it as follows:
\be
\label{15}
 R_{\mu_1}\equiv
\frac{\la \vec{k}_{\perp}^{4}\ra  }{\la \vec{k}_{\perp}^{2}\ra^2}
 \sim \frac{ \la g^2G_{\mu\nu}^aG_{\mu\nu}^a\ra\la \bar{q}q\ra^2}
{\la \bar{q}ig\sigma_{\mu\nu}
 G_{\mu\nu}^a\frac{\lambda^a}{2} q \ra^2}
\sim(\frac{\alpha_s(\mu_1)}{\alpha_s(\mu_2)})^{\frac{-20}{3b}}
\cdot R_{\mu_2}
 \ee

Let us summarize the results of the section.
The constraints ($\bullet 1 -\bullet 3$) have very general origin
and should be fulfilled in any phenomenological description based on QCD.
The numerical constraints  ($\bullet 4 -\bullet 6$)
have much less generality because they have been obtained
from QCD sum rules with inevitable for this method approximations.
All numerical results obtained from QCD sum rules
are normalized at $\mu^2\sim 1GeV^2$. At the same time,
the model $wf$ we are going to construct
should be normalized at the lowest possible point which
is about $\mu^2\sim 0.25GeV^2$. We shall use the
anomalous dimensions shown above in order to evaluate
all numerical constraints at the  lowest normalization point.

 \vspace {0.3cm}
{\bf 3. The model wave function.}
 \vspace {0.3cm}

Let us start our discussion  from the analysis of the $wf$
motivated by constituent quark model \cite{Cotanch}, \cite{Isgur1}-
\cite{BHL} (CQM)
\footnote{Here
we neglect all terms
in QCM related to spin part of constituents. In particular,
we do  not consider Melosh transformation and other ingredients
of the light cone$\Leftrightarrow$equal time connection. It does
not effect   qualitative  results presented in the next section.}.
Such a function is known to give a reasonable description
of static hadron  properties.
The Brodsky-Huang-Lepage prescription \cite{BHL}
 leads to the following form for the pion $wf$:
 \be
\label{16}
 \psi(\k , x)_{CQM}=A
\exp(-\frac{\k+m^2}{8\beta^2 x(1-x)} ),
 \ee
We call this function  as the constituent quark model
 $wf$. It  satisfies
two constraints ($\bullet 2, \bullet 3$), but not   ($\bullet 1$)
because of the nonzero magnitude for the constituent mass $m$.
We take the standard set for QCD parameters:
\be
\label{17}
 \alpha_s(1GeV)=0.34,~~~\Lambda_{QCD}=200 MeV.
 \ee
The lowest possible normalization point $\mu_0$ is defined
as the place where $\alpha_s(\mu_0)\simeq 0.7$. This corresponds to
the    QCD sum rules analysis  \cite{Shif1}, where "almost"
renormalization invariant combination $\alpha_s\la\bar{q}q\ra^2$
comes into the game
and it is numerically well known. At the same time  the chiral condensate
$\la\bar{q}q\ra$  at the  lowest possible
normalization point is known from PCAC.

We made the standard choice for the parameter $m\simeq 330 MeV$
in accordance with   its physical meaning.   The parameter
$\beta$ is determined from the numerical constraint ($\bullet 4$)
for the mean   value $\la\k\ra$ \footnote{Do not
confuse our parameter $\la\k\ra$ which is well defined
in terms of QCD matrix element (\ref{6}) with the one which is defined
in terms of CQM as follows:
$$\la\k\ra_{q\bar{q}}\equiv\frac{\int d^2\k dx|\psi(\k , x)_{CQM}|^2\k}
{\int d^2\k dx|\psi(\k , x)_{CQM}|^2 } $$
and should be extracted from somewhere else.
Numerically they are not very different, but we prefer
to use one and the same procedure to specify parameters
for all wave functions.}.
 Parameter $A$
is determined by the normalization eq.(\ref{9}).
As we mentioned earlier, we have to renormalize all   moments
to the lowest possible point to model $wf$.
In our case the evaluation of $\la\k\ra$ is defined by eq.(\ref{11}).
With our set of parameters (\ref{17})
  and ($\bullet 5$) we have the following
mean value for $\k$ at lowest normalization point:
\be
\label{18}
\la\k\ra_{\mu_0}=0.14\,GeV^2,
\ee
which will be used through this paper. This number corresponds
to $\beta\simeq0.3GeV $ which is within reasonable parametric
region.
 To make function    wider in the longitudinal  direction
  (constraint $\bullet 4$) one can insert to formula (\ref{16})
the additional factor
\be
\label{19}
(1+g(\mu)[(2x-1)^2-\frac{1}{5}]),~~g(\mu_1)
=g(\mu_2)\cdot(\frac{\alpha_s(\mu_1)}{\alpha_s(\mu_2)})^{50/9b}
\ee
with additional parameter $g(\mu)$.
With this new parameter $g(\mu)$ one can  adjust $\la\xi^2\ra$
as appropriate. For the asymptotic distribution
amplitude   parameter
$g=0$.

We are ready now to discuss the model $wf$ in QCD.
  Before to design  $\psi_{QCD}(\k,x)$,
let us explain  what do we mean by that.
We define the {\it nonperturbative wave function} $\psi(\k,x, \mu)_{QCD} $
through its moments which can be expressed in terms of the nonperturbative
matrix elements (\ref{d}). As is known, all nonperturbative matrix elements
are defined in such a way that all gluon's and quark's
virtualities smaller than some parameter $\mu$ (point of normalization)
are hidden in the definition of the " nonperturbative
matrix elements". All virtualities larger than that should be
take into account perturbatively. In particular, all perturbative tails
like $1/\k$ should be subtracted from the definition
of the nonperturbative $wf$.
 The same procedure should be applied
for the  calculation of nonperturbative vacuum condensate $\la
G_{\mu\nu}^2\ra$,
where the perturbative part related to free gluon propagator
$1/k^2$ should be subtracted.

With these general remarks in mind
we propose the following form for the
  nonperturbative wave function $\psi(\k,x, \mu_0)_{QCD} $
at the lowest normalization point:
 \be
\label{20}
 \psi(\k , x, \mu_0)_{QCD}=A
\exp(-\frac{\k }{8\beta^2 x(1-x)} )\cdot\{1+g(\mu_0)[(2x-1)^2-\frac{1}{5}]\}.
 \ee
In comparison with the constituent
quark model the "only" difference is the absence
on the mass term $\sim m$ in the exponent. As we shall see
in the next section it does make a big difference.
This function satisfies  all fundamental constraints
($\bullet 1-\bullet 3$). The dimensional  parameters can be
determined from the numerical
relations ($\bullet 4-\bullet 5$) in the same way as before.
Parameter $\beta$ for this parametrization is found to be
$\beta\simeq$ 0.3 GeV (it corresponds to $R=2.2$ and $\la\k\ra=0.14 GeV^2$)

We would like to emphasize that the nonzero mass in $\psi(\k , x)_{CQM}$
 was  unavoidable part of the wave function.
  We do not see any room for such    term in QCD,  because
its presence   would mean
 the following behavior of the
large moments in longitudinal direction:
   \be
\label{21}
 \la\xi^n\ra=\int_{-1}^1d\xi \xi^n\phi(\xi)
\sim \int_{-1}^1d\xi \xi^n\exp(-\frac{1}{1-\xi^2})\sim
\exp(-\sqrt{n}), ~n\rightarrow\infty.
  \ee
It is in contradiction to  $1/n^2$ behavior
(\ref{5}), ($\bullet 1$). Let us stress that such a behavior
is the result of the calculation of the correlation function
(\ref{2}). We do not see any possibilities to change this behavior
from $1/n^2$ to $ \exp(-\sqrt{n})$.
 If such a behavior were occurred, it
would mean that    the
strong cancellation between one-loop diagram (\ref{3})
and higher loop corrections $\sim \alpha_s^k$
  takes place.
Such a cancellation looks even less probable,
if one takes into account that the aforementioned cancellation
must take place for each given number $n$ at large $n$.
We do not believe that it might happen in QCD.

Up to now we did not discuss the influence of
 our last ``numerical" constraint
($\bullet 6$) denoted by $R$. Large number for this
parameter means a  noticeable fluctuations of the momentum
in transverse direction. To satisfy this constraint
we need to spread out the $wf$ to make it wider.
It can be done in arbitrary way. In particular, one may try
to put one more hump apart from the main Gaussian term
described by eq.(\ref{20}). The only requirement is:
it has to fall off  fast enough at large $\k$.

With these remarks in mind we suggest the following
QCD motivated $wf$ (we call it as $\psi(\k , x, \mu_0)_{QCD+}$)
which can be adjusted to satisfy all
six requirements mentioned in the previous section:
 \be
\label{22}
 \psi(\k , x, \mu_0)_{QCD+}=A
\{e^{-\frac{\k }{8\beta^2 x(1-x)} }+
c\cdot e^{-(\frac{\k }{8\beta^2 x(1-x)}-l)^2 } \}\cdot
\ee
\be
 \{1+g(\mu)[(2x-1)^2-\frac{1}{5}]\}.  \nonumber
 \ee
The physical meaning of the parameters $c, l$ is clear.
Parameter $c$ determines the magnitude of the second
hump and  parameter $l$ describes its  distance
from  the main term. As we shall see in order to
match parameters $l,c$ with the calculated
ratio $R$ we need to have the magnitude
of the second hump about $1/10$.
The maximum of the second hump is located in $\k\sim 0.8 GeV^2$ region.
To be more specific,
we renormalized the parameter $R$ found from QCD to the lowest
normalization point according to formula (\ref{15}).
It is found to be
\be
\label{23}
R(\mu_0)\simeq 3\div 4
\ee
 We display this function with parameters $\beta=0.15$ GeV
, c = 0.15, l = 30 (which correspond
to $R\simeq 4, \la\k\ra =0.14 GeV^2$)
on Fig.1 at the central point $x=1/2$.

Let us summarize.
We constructed three wave functions.
The first one, $\psi_{CQM}$ is motivated by
quark model with its specific mass parameters.
Two other models are motivated by QCD consideration.
Despite of this difference,
all these models have Gaussian behavior at large $\k$.
However, in the case of $\psi_{CQM}$ this behavior is
related to the nonrelativistic ocsilator model,
while for QCD motivated models this behavior
is provided by  constraints discussed in the previous section.

Contrary to the CQM, the QCD motivated wave functions
do not contain the mass parameter $m\simeq 300 MeV$ which is an essential
ingredient of any quark model. Such a term
is absolutely forbidden from the QCD point of view.

The difference between $\psi_{QCD}$ and $\psi_{QCD+}$
is not fundamental, but rather quantitative.
Nevertheless, we believe that it is worth to mention
some new effects (noticeable fluctuations of
transverse momentum) which the function
$\psi_{QCD+}$   brings. The broadening
in  the transverse direction is the main difference between
functions
$\psi_{QCD}$ and $\psi_{QCD+}$.
In the next section we   discuss
the contribution to the pion form factor
caused by these three wave functions.
We shall see the qualitative difference in behavior
on $Q^2$, which is our main point.

\vspace {0.3cm}
 {\bf 4. Pion Form Factor.}
 \vspace {0.3cm}

The starting point is the famous Drell-Yan formula
\cite{DY} (for modern, QCD- motivated  employing of this formula,
see \cite{Brod1}),
where the $F_{\pi}(Q^2)$ is expressed in terms of full
wave functions:
\be
\label{DY}
F_{\pi}(Q^2)=\int\frac{dx d^2\vec{k}_{\perp}}
{16\pi^3}\Psi^*_{BL}
(x ,\vec{k}_{\perp}+(1-x) \vec{q}_{\perp})
\Psi_{BL}(x ,\vec{k}_{\perp}),
\ee
where $q^2=-\vec{q}_{\perp}^{~}2=-Q^2$ is the momentum transfer.
In this formula, the $\Psi_{BL}(x ,\vec{k}_{\perp})$ is
 the full wave function;
the perturbative tail of $\Psi_{BL}(x ,\vec{k}_{\perp})$ behaves as
$\alpha_s/ \k$ for large $\k$ and should be taken into account
explicitly in the calculations. This gives the
 one-gluon-exchange
(asymptotically leading) formula for the form factor in
terms of distribution amplitude $\phi(x)$ \cite{Brod}.

Below is the QCD motivated interpretation of this formula.
Let us remind, that   the formula (\ref{DY})  takes into account
only the valence Fock states.  The formula would be exact
if all Fock states were taken into account.
Besides that, $\k$ in this formula was
originally thought to be  the usual (not covariant)
 perpendicular momentum of the constituents, and not the
mean value $\la\k\ra$   defined in QCD, as a gauge invariant object.
However we make the
{\bf assumption} that   it is one and the same
variable. The physics behind of it
  can be explained in the following way.

In the formula (\ref{DY})   we effectively take into account
 some gluons (not all of them), which inevitably
 are participants of our definition of $wf$. These gluons mainly carry the
transverse momentum (which anyhow, does not exceed QCD scale
of order $ \sim 1 GeV$) and/or small amount of  the longitudinal momentum.
The contributions of the
gluons carrying the {\it finite}
longitudinal momentum fraction  are neglected in (\ref{DY}).
This is the main assumption.
 It can be justified
by the direct calculation  \cite{Cher}
   of quark-antiquark-gluon
(with
 finite momentum fraction)
contribution to $\pi$ meson form factor at large $Q^2$ within the
standard technique of the operator product expansion.
By technical reasons
the corresponding calculation has not been  completed, however
it was found
that the characteristic scale which enters into the game is
of order $1 GeV^2$. Thus, it is very unlikely
 that these contributions
can  be important at $Q^2\gg 1 GeV^2$. The second calculation,
which confirms this point, comes from the light cone QCD sum
rules \cite{Braun1}. This is almost model independent calculation
demonstrates that the quark-antiquark-gluon
(with
 finite momentum fraction)
contribution does
not exceed $ 20\% $ at available $Q^2$.

Thus, we expect, that by taking into account
the only "soft" gluon contribution  (hidden
in the definition of $\k$ (\ref{1})), we catch  the main effect.
Again, there is no proof for that within QCD,
and the only argumentation which can be delivered now in favor of it,
is based on the intuitive picture of quark model, where
current quark and  soft gluons form a constituent quark
 with original quantum numbers.
No evidence of the gluon playing the role of a valent participant
with a finite amount of momentum, is found.

{}From the viewpoint of the operator product expansion, the
assumption formulated above, corresponds to the {\bf summing up} a
subset of higher-dimension power corrections. This subset
actually is formed from the infinite number of soft gluons
and  unambiguously
singled out by the definition of nonperturbative $wf$ (\ref{1}).

In the following, we preserve the notation
$\Psi_{BL}(x,\vec{k}_{\perp})$ for
the nonperturbative, soft part  only. It should not confuse
the reader.

The formula (\ref{DY}) is written
in terms of Brodsky and Lepage notations \cite{Brod1};
 the relation  to our wave function $\psi( \vec{k}_{\perp}, x)$
is given by formula (\ref{9}).

With these general remarks in mind, we would like
to present the results of calculation, based on three wave functions
discussed in the previous section.

The first calculation, based on
$\psi_{CQM}(\k, x) $ is the standard one. The
analogous calculations with ocsilator-like
$wf$  have been done many times with many additional improvements,
see i.e. \cite{Cotanch}.
One can fit the dimensional parameters in such a way
that the description at low $Q^2$ will be perfect.
However, our goal is different
and   we are interested in the behavior at large enough $Q^2\gg 1GeV^2$.
  We expect that described here approach makes sense
only at high enough $Q^2$. We display the corresponding behavior
as a curve 1 on Fig.2.

The main feature of this behavior --
it gives very reasonable
magnitude for the intermediate region about few $GeV^2$
and it
starts to fall off right after that.
We expect that {\bf any} reasonable, well localized,
based on quark model wave function
with the scale $\sim \la\k\ra\sim m^2 $ leads to the similar
behavior.  Let me stress: we are not pretending to have made
a reliable calculation of  the form factor here;
we displayed this contribution only for the
{\bf illustrative}
purposes.

 Currently, much more interesting for us is
the calculation, based on QCD motivated models.
 We display the corresponding contribution to the
$Q^2F_{\pi}(Q^2)$  for the $\psi_{QCD}(x,\k)$
on Fig. 2 as curve 2. The {\bf qualitative}
difference between this curve and the previous one (curve 1),
 is much slower fall  off at large $Q^2$
for the model $wf$ $\psi_{QCD}$.
The qualitative reason for that is the absence of the
mass term in $\psi_{QCD}$, see discussion after the formula (\ref{21}).
Precisely this term
was responsible for the very steep behavior in the
previous calculation  with quark model wave function.

 The declining of the form factor getting even slower
if one takes into account the property of the broadening
of  $wf$ in transverse direction. The corresponding
 contribution   based on $\psi_{QCD+}$ is displayed
 on Fig.2 as curve 3. We notice an additional
slowing down  of the declining of the magnitude $Q^2F(Q^2)$
in the intermediate region of $Q^2$.
The explanation of this effect is the following.
When $Q^2$ is getting bigger and bigger, the contribution coming
from the overlap of   two humps ($\psi_{QCD+}  $
at $x=1/2$ is displayed
on Fig.1)
 starts to grow. These humps are well separated
(in order to satisfy
constraint ($\bullet 6$)) from each other by the value
of order $1GeV^2$. Therefore, we expect that this contribution starts
to grow at high enough $Q^2\gg 4GeV^2$. As the result, the curve 3
looks more horizontal than the previous one. It is getting lower
  because the general scale has been changed
when we passed from $\psi_{QCD}$ to $\psi_{QCD+}$.
  Let us remind that we keep
$\la\k\ra$ fixed ($\bullet 5$) in all cases.
It leads to some changes of  dimensional parameters
because the small hump in the $\psi_{QCD+}$
gives a noticeable contribution to $\la\k\ra$
in spite of the fact that it comes with very small relative weight
(coefficient $c\sim 1/10$).

Our last qualitative remark
is some note
that the results strongly depend
on parameter $\la\xi^2\ra$.    We displayed
on Fig.3 the same three curves as on Fig.2 with the only difference
in coefficient $g(\mu)$ (\ref{19}, \ref{20}, \ref{22})
   We set   $g(\mu_0)=2$ which corresponds to
$\la\xi^2\ra=0.3$. It makes a $wf$ wider in
longitudinal direction. The soft contribution getting bigger
when a wider (in longitudinal direction) wave function is used.
The same effect was observed
in the  recent calculation \cite{Braun1}, where
a quite different method has been used.

 The  contribution
under consideration is subject to Sudakov corrections.
An estimate of these corrections reveals that they are small enough
in this intermediate $Q^2$ region. Besides that,
we will be on the   safe side if we say that the hard
(leading twist ) contribution to $Q^2F(Q^2)\simeq 0.2 GeV^2$
\cite{Kroll}, \cite{Kiss}, \cite{Braun1}. It should be
added to the soft terms displayed on Fig.2, Fig.3.

The precise fitting of the pion form factor was not   among the goals of
this paper. Rather, we wanted to demonstrate how the qualitative
properties of a nonperturbative $wf$, derived from the
QCD analysis might  significantly change its behavior.

\vspace {0.3cm}
 {\bf 5. Summary and Outlook.}
 \vspace {0.3cm}

The main goal of the present paper was the analysis
of the nonperturbative $wf$ from QCD point of view.
We found qualitatively  different results in comparison  with the
wave functions motivated by  quark model. We believe that this difference
is responsible for the
qualitative explanation of
     dimensional
counting rules which work well even at very modest energies.

 The standard point of view for
the phenomenological success of the dimensional
counting rules is the
   predigest  that the leading twist contribution
plays the main role in most cases.
We suggest here some different explanation for this
phenomenological success.
Our explanation of  the   slow falling off  of the soft contribution
with energy is due to the
specific properties of   nonperturbative $wf$.
 In particular,
we found a {\bf new} scale ($\sim 1GeV^2$) in the
 problem, in addition to the standard low energy parameter
$\la\k\ra\simeq 0.1 GeV^2$.

Our next remark can be formulated as follows.
Exclusive, as well as inclusive amplitudes can be expressed
in terms   of the {\bf one and the same} particular hadron
$wf$. Therefore, if our explanation
(related to specific form of $\psi(\k,x)$) of a temporary simulation of
 the leading twist behavior  is considered as a reasonable one , then:

1. in the analysis of inclusive amplitudes
one may expect  the same effect ( it is our conjecture );

2. one may try to implement the intrinsic transverse
momentum dependence into the inclusive calculations.

In particular, one may try to use the following prescription
for the $\pi$ meson distribution function
(and anologously for nucleon) at $x \Rightarrow 1$:
\be
\label{*}
G_{q/\pi}(x,Q^2)
\Rightarrow \{\int\frac{d^2k_{\perp}}{x(1-x)}\exp(- \frac{\k}{x(1-x)})
\} G_{q/\pi}(x,Q^2)
\ee
The analogous formula (without $x$ dependence in the exponent)
has been suggested many years ago \cite{Parisi}, see also
\cite{Cox}.
Our remark is that this $x$ dependence is essential point and
should be introduced to the formula to satisfy our constraints.
In terms of \cite{Parisi}, \cite{Cox} it corresponds to the
non-universality of their ``constant" $\la\k\ra$ which now will depend on $x$.

To support this conjecture, we would like to mention few inclusive processes
where the intrinsic
transverse distribution might be essential. First of all, it is
Drell-Yan amplitude $\pi+N\rightarrow\mu^-\mu^++X$
which is parametrized as follows (for
references and recent development see \cite{Khoze1}):
\be
\label{angle}
\frac{1}{\sigma}\frac{d\sigma}{d\Omega}\sim
1+\lambda\cos^2\theta+\mu\sin2\theta\cos\phi+\frac{\nu}{2}\sin^2\theta
\cos2\phi .
\ee
Here $\theta,\phi$ are angles defined in the muon pair rest frame and
$\lambda, \mu, \nu$ are coefficients.
In the naive parton model the coefficient are $\lambda=1,
\mu=\nu=0$. Experimental results do not support
this naive prediction. Recently, some improvements have been made
\cite{Khoze1}, but some problems are still remain. In particular,
the Lam-Tung sum rule \cite{Tung1}, $1-\lambda-2\nu=0$
is violated by experimental data and the improved model \cite{Khoze1}
still can not   explain the behavior
$1-\lambda-2\nu $ as a function of $Q_{\perp}^2$ (
$Q_{\perp}^2$ is the transverse
momentum
of the lepton pair).

Due to the fact that $Q^2$ is not large enough in this
experiment one may expect
that the intrinsic distribution (\ref{*}) might be essential.

Analogously, we would expect that the
$\pi$ meson structure (\ref{*}) may effect the analysis
of azimuthal asymmetries in semi-exclusive
amplitudes like
$l+p\rightarrow l'+h+X$. For references and   recent development see
\cite{Khoze2}, where intrinsic $\k$ has been introduced
in the standard way without $x$ dependence in the exponent.

One may find many examples like that where the standard
parton picture does not work well. We would like to mention
here the recent analysis \cite{Tung2}
of the direct photon production
($\pi+p\rightarrow\gamma + X$), with the result
that perturbative QCD can not explain the data. Some nonperturbative
broadening factor in transverse direction should be implemented.
One may hope that formula (\ref{*})
may improve the agreement with experiment.

\vspace {0.3cm}
 {\bf 5. Acknowledgments.}
 \vspace {0.3cm}

 We wish to thank Stan Brodsky for useful critical
comments and discussions. We also thankful to
 Kent Hornbostel for his help in  numerical
 computations.

 This work is supported by the Texas National Research
Laboratory Commission under  grant \# RCFY 93-229.

\newpage
{\bf FIGURE CAPTIONS}\\[2.0cm]
 {\bf Fig.1 } QCD Wavefunction ( unnormalized ) $\psi(\k, x, \mu_0)_{QCD+}$
versus transverse momentum $ k_{\perp} $.
\vspace{0.3cm}\\
{\bf Fig.2 } Pion Form Factor  $Q^2\cdot F_{\pi}(Q^2)$ versus $Q^2$.
 Line 1 corresponds to the $\psi(\k , x)_{CQM}$,
with parametres, $R\simeq 2.0$  (\ref {23}),
 $\la\k\ra =0.14$ $ GeV^2$   (\ref{6}) ,
 line 2  corresponds to the
$\psi(\k , x, \mu_0)_{QCD} $
with parametres, $R\simeq 2.2,
\la\k\ra$ =0.14 $GeV^2,
\la\xi^2\ra$ =0.2,
and line 3 shows the result for two humped
 ( in transverse direction )
wavefunction $\psi(\k , x,\mu_0)_{QCD+}$,
 with  parameters  $R\simeq 4.0,
 \la\k\ra $ =0.14 $GeV^2,
 \la\xi^2\ra$ =0.2
\vspace{0.3cm}\\
{\bf Fig.2 } The same Pion Form Factor for $\la\xi^2\ra=0.3$\\[2.0cm]

{\bf FIGURES }\\[1.0cm]

\vspace{0.3cm}

\setlength{\unitlength}{0.240900pt}
\ifx\plotpoint\undefined\newsavebox{\plotpoint}\fi
\begin{picture}(1500,900)(0,0)
\font\gnuplot=cmr10 at 10pt
\gnuplot
\sbox{\plotpoint}{\rule[-0.200pt]{0.400pt}{0.400pt}}%
\put(176.0,68.0){\rule[-0.200pt]{303.534pt}{0.400pt}}
\put(176.0,68.0){\rule[-0.200pt]{0.400pt}{194.888pt}}
\put(176.0,68.0){\rule[-0.200pt]{4.818pt}{0.400pt}}
\put(154,68){\makebox(0,0)[r]{$0$}}
\put(1416.0,68.0){\rule[-0.200pt]{4.818pt}{0.400pt}}
\put(176.0,230.0){\rule[-0.200pt]{4.818pt}{0.400pt}}
\put(154,230){\makebox(0,0)[r]{$0.2$}}
\put(1416.0,230.0){\rule[-0.200pt]{4.818pt}{0.400pt}}
\put(176.0,392.0){\rule[-0.200pt]{4.818pt}{0.400pt}}
\put(154,392){\makebox(0,0)[r]{$0.4$}}
\put(1416.0,392.0){\rule[-0.200pt]{4.818pt}{0.400pt}}
\put(176.0,553.0){\rule[-0.200pt]{4.818pt}{0.400pt}}
\put(154,553){\makebox(0,0)[r]{$0.6$}}
\put(1416.0,553.0){\rule[-0.200pt]{4.818pt}{0.400pt}}
\put(176.0,715.0){\rule[-0.200pt]{4.818pt}{0.400pt}}
\put(154,715){\makebox(0,0)[r]{$0.8$}}
\put(1416.0,715.0){\rule[-0.200pt]{4.818pt}{0.400pt}}
\put(176.0,877.0){\rule[-0.200pt]{4.818pt}{0.400pt}}
\put(154,877){\makebox(0,0)[r]{$1$}}
\put(1416.0,877.0){\rule[-0.200pt]{4.818pt}{0.400pt}}
\put(176.0,68.0){\rule[-0.200pt]{0.400pt}{4.818pt}}
\put(176,23){\makebox(0,0){$0$}}
\put(176.0,857.0){\rule[-0.200pt]{0.400pt}{4.818pt}}
\put(428.0,68.0){\rule[-0.200pt]{0.400pt}{4.818pt}}
\put(428,23){\makebox(0,0){$0.2$}}
\put(428.0,857.0){\rule[-0.200pt]{0.400pt}{4.818pt}}
\put(680.0,68.0){\rule[-0.200pt]{0.400pt}{4.818pt}}
\put(680,23){\makebox(0,0){$0.4$}}
\put(680.0,857.0){\rule[-0.200pt]{0.400pt}{4.818pt}}
\put(932.0,68.0){\rule[-0.200pt]{0.400pt}{4.818pt}}
\put(932,23){\makebox(0,0){$0.6$}}
\put(932.0,857.0){\rule[-0.200pt]{0.400pt}{4.818pt}}
\put(1184.0,68.0){\rule[-0.200pt]{0.400pt}{4.818pt}}
\put(1184,23){\makebox(0,0){$0.8$}}
\put(1184.0,857.0){\rule[-0.200pt]{0.400pt}{4.818pt}}
\put(1436.0,68.0){\rule[-0.200pt]{0.400pt}{4.818pt}}
\put(1436,23){\makebox(0,0){$1$}}
\put(1436.0,857.0){\rule[-0.200pt]{0.400pt}{4.818pt}}
\put(176.0,68.0){\rule[-0.200pt]{303.534pt}{0.400pt}}
\put(1436.0,68.0){\rule[-0.200pt]{0.400pt}{194.888pt}}
\put(176.0,877.0){\rule[-0.200pt]{303.534pt}{0.400pt}}
\put(176.0,68.0){\rule[-0.200pt]{0.400pt}{194.888pt}}
\put(507,68){\usebox{\plotpoint}}
\multiput(1220.00,68.61)(2.472,0.447){3}{\rule{1.700pt}{0.108pt}}
\multiput(1220.00,67.17)(8.472,3.000){2}{\rule{0.850pt}{0.400pt}}
\multiput(1232.58,71.00)(0.493,1.012){23}{\rule{0.119pt}{0.900pt}}
\multiput(1231.17,71.00)(13.000,24.132){2}{\rule{0.400pt}{0.450pt}}
\multiput(1245.58,97.00)(0.493,2.519){23}{\rule{0.119pt}{2.069pt}}
\multiput(1244.17,97.00)(13.000,59.705){2}{\rule{0.400pt}{1.035pt}}
\multiput(1258.58,161.00)(0.493,0.933){23}{\rule{0.119pt}{0.838pt}}
\multiput(1257.17,161.00)(13.000,22.260){2}{\rule{0.400pt}{0.419pt}}
\multiput(1271.58,176.01)(0.492,-2.650){21}{\rule{0.119pt}{2.167pt}}
\multiput(1270.17,180.50)(12.000,-57.503){2}{\rule{0.400pt}{1.083pt}}
\multiput(1283.58,116.71)(0.493,-1.805){23}{\rule{0.119pt}{1.515pt}}
\multiput(1282.17,119.85)(13.000,-42.855){2}{\rule{0.400pt}{0.758pt}}
\multiput(1296.00,75.93)(0.824,-0.488){13}{\rule{0.750pt}{0.117pt}}
\multiput(1296.00,76.17)(11.443,-8.000){2}{\rule{0.375pt}{0.400pt}}
\put(1309,67.67){\rule{2.891pt}{0.400pt}}
\multiput(1309.00,68.17)(6.000,-1.000){2}{\rule{1.445pt}{0.400pt}}
\put(507.0,68.0){\rule[-0.200pt]{171.762pt}{0.400pt}}
\put(1321.0,68.0){\rule[-0.200pt]{27.703pt}{0.400pt}}
\put(176,877){\usebox{\plotpoint}}
\multiput(176.00,875.95)(2.695,-0.447){3}{\rule{1.833pt}{0.108pt}}
\multiput(176.00,876.17)(9.195,-3.000){2}{\rule{0.917pt}{0.400pt}}
\multiput(189.00,872.92)(0.600,-0.491){17}{\rule{0.580pt}{0.118pt}}
\multiput(189.00,873.17)(10.796,-10.000){2}{\rule{0.290pt}{0.400pt}}
\multiput(201.58,861.54)(0.493,-0.616){23}{\rule{0.119pt}{0.592pt}}
\multiput(200.17,862.77)(13.000,-14.771){2}{\rule{0.400pt}{0.296pt}}
\multiput(214.58,844.77)(0.493,-0.853){23}{\rule{0.119pt}{0.777pt}}
\multiput(213.17,846.39)(13.000,-20.387){2}{\rule{0.400pt}{0.388pt}}
\multiput(227.58,822.14)(0.493,-1.052){23}{\rule{0.119pt}{0.931pt}}
\multiput(226.17,824.07)(13.000,-25.068){2}{\rule{0.400pt}{0.465pt}}
\multiput(240.58,794.02)(0.492,-1.401){21}{\rule{0.119pt}{1.200pt}}
\multiput(239.17,796.51)(12.000,-30.509){2}{\rule{0.400pt}{0.600pt}}
\multiput(252.58,760.99)(0.493,-1.408){23}{\rule{0.119pt}{1.208pt}}
\multiput(251.17,763.49)(13.000,-33.493){2}{\rule{0.400pt}{0.604pt}}
\multiput(265.58,724.60)(0.493,-1.527){23}{\rule{0.119pt}{1.300pt}}
\multiput(264.17,727.30)(13.000,-36.302){2}{\rule{0.400pt}{0.650pt}}
\multiput(278.58,685.22)(0.493,-1.646){23}{\rule{0.119pt}{1.392pt}}
\multiput(277.17,688.11)(13.000,-39.110){2}{\rule{0.400pt}{0.696pt}}
\multiput(291.58,642.64)(0.492,-1.832){21}{\rule{0.119pt}{1.533pt}}
\multiput(290.17,645.82)(12.000,-39.817){2}{\rule{0.400pt}{0.767pt}}
\multiput(303.58,599.96)(0.493,-1.726){23}{\rule{0.119pt}{1.454pt}}
\multiput(302.17,602.98)(13.000,-40.982){2}{\rule{0.400pt}{0.727pt}}
\multiput(316.58,555.84)(0.493,-1.765){23}{\rule{0.119pt}{1.485pt}}
\multiput(315.17,558.92)(13.000,-41.919){2}{\rule{0.400pt}{0.742pt}}
\multiput(329.58,510.63)(0.492,-1.832){21}{\rule{0.119pt}{1.533pt}}
\multiput(328.17,513.82)(12.000,-39.817){2}{\rule{0.400pt}{0.767pt}}
\multiput(341.58,468.22)(0.493,-1.646){23}{\rule{0.119pt}{1.392pt}}
\multiput(340.17,471.11)(13.000,-39.110){2}{\rule{0.400pt}{0.696pt}}
\multiput(354.58,426.35)(0.493,-1.607){23}{\rule{0.119pt}{1.362pt}}
\multiput(353.17,429.17)(13.000,-38.174){2}{\rule{0.400pt}{0.681pt}}
\multiput(367.58,385.73)(0.493,-1.488){23}{\rule{0.119pt}{1.269pt}}
\multiput(366.17,388.37)(13.000,-35.366){2}{\rule{0.400pt}{0.635pt}}
\multiput(380.58,347.60)(0.492,-1.530){21}{\rule{0.119pt}{1.300pt}}
\multiput(379.17,350.30)(12.000,-33.302){2}{\rule{0.400pt}{0.650pt}}
\multiput(392.58,312.37)(0.493,-1.290){23}{\rule{0.119pt}{1.115pt}}
\multiput(391.17,314.68)(13.000,-30.685){2}{\rule{0.400pt}{0.558pt}}
\multiput(405.58,279.63)(0.493,-1.210){23}{\rule{0.119pt}{1.054pt}}
\multiput(404.17,281.81)(13.000,-28.813){2}{\rule{0.400pt}{0.527pt}}
\multiput(418.58,249.14)(0.493,-1.052){23}{\rule{0.119pt}{0.931pt}}
\multiput(417.17,251.07)(13.000,-25.068){2}{\rule{0.400pt}{0.465pt}}
\multiput(431.58,222.26)(0.492,-1.013){21}{\rule{0.119pt}{0.900pt}}
\multiput(430.17,224.13)(12.000,-22.132){2}{\rule{0.400pt}{0.450pt}}
\multiput(443.58,198.77)(0.493,-0.853){23}{\rule{0.119pt}{0.777pt}}
\multiput(442.17,200.39)(13.000,-20.387){2}{\rule{0.400pt}{0.388pt}}
\multiput(456.58,177.16)(0.493,-0.734){23}{\rule{0.119pt}{0.685pt}}
\multiput(455.17,178.58)(13.000,-17.579){2}{\rule{0.400pt}{0.342pt}}
\multiput(469.58,158.37)(0.492,-0.669){21}{\rule{0.119pt}{0.633pt}}
\multiput(468.17,159.69)(12.000,-14.685){2}{\rule{0.400pt}{0.317pt}}
\multiput(481.58,142.80)(0.493,-0.536){23}{\rule{0.119pt}{0.531pt}}
\multiput(480.17,143.90)(13.000,-12.898){2}{\rule{0.400pt}{0.265pt}}
\multiput(494.00,129.92)(0.539,-0.492){21}{\rule{0.533pt}{0.119pt}}
\multiput(494.00,130.17)(11.893,-12.000){2}{\rule{0.267pt}{0.400pt}}
\multiput(507.00,117.92)(0.652,-0.491){17}{\rule{0.620pt}{0.118pt}}
\multiput(507.00,118.17)(11.713,-10.000){2}{\rule{0.310pt}{0.400pt}}
\multiput(520.00,107.93)(0.758,-0.488){13}{\rule{0.700pt}{0.117pt}}
\multiput(520.00,108.17)(10.547,-8.000){2}{\rule{0.350pt}{0.400pt}}
\multiput(532.00,99.93)(0.950,-0.485){11}{\rule{0.843pt}{0.117pt}}
\multiput(532.00,100.17)(11.251,-7.000){2}{\rule{0.421pt}{0.400pt}}
\multiput(545.00,92.93)(1.378,-0.477){7}{\rule{1.140pt}{0.115pt}}
\multiput(545.00,93.17)(10.634,-5.000){2}{\rule{0.570pt}{0.400pt}}
\multiput(558.00,87.93)(1.378,-0.477){7}{\rule{1.140pt}{0.115pt}}
\multiput(558.00,88.17)(10.634,-5.000){2}{\rule{0.570pt}{0.400pt}}
\multiput(571.00,82.94)(1.651,-0.468){5}{\rule{1.300pt}{0.113pt}}
\multiput(571.00,83.17)(9.302,-4.000){2}{\rule{0.650pt}{0.400pt}}
\put(583,78.17){\rule{2.700pt}{0.400pt}}
\multiput(583.00,79.17)(7.396,-2.000){2}{\rule{1.350pt}{0.400pt}}
\multiput(596.00,76.95)(2.695,-0.447){3}{\rule{1.833pt}{0.108pt}}
\multiput(596.00,77.17)(9.195,-3.000){2}{\rule{0.917pt}{0.400pt}}
\put(609,73.17){\rule{2.500pt}{0.400pt}}
\multiput(609.00,74.17)(6.811,-2.000){2}{\rule{1.250pt}{0.400pt}}
\put(621,71.67){\rule{3.132pt}{0.400pt}}
\multiput(621.00,72.17)(6.500,-1.000){2}{\rule{1.566pt}{0.400pt}}
\put(634,70.67){\rule{3.132pt}{0.400pt}}
\multiput(634.00,71.17)(6.500,-1.000){2}{\rule{1.566pt}{0.400pt}}
\put(647,69.67){\rule{3.132pt}{0.400pt}}
\multiput(647.00,70.17)(6.500,-1.000){2}{\rule{1.566pt}{0.400pt}}
\put(672,68.67){\rule{3.132pt}{0.400pt}}
\multiput(672.00,69.17)(6.500,-1.000){2}{\rule{1.566pt}{0.400pt}}
\put(660.0,70.0){\rule[-0.200pt]{2.891pt}{0.400pt}}
\put(711,67.67){\rule{2.891pt}{0.400pt}}
\multiput(711.00,68.17)(6.000,-1.000){2}{\rule{1.445pt}{0.400pt}}
\put(685.0,69.0){\rule[-0.200pt]{6.263pt}{0.400pt}}
\put(723.0,68.0){\rule[-0.200pt]{171.762pt}{0.400pt}}
\end{picture}\\[0.3cm]
$${\bf Fig. 1}$$

\newpage

$${\bf Q^2F(Q^2)~~ vs ~~ Q^2}$$

\setlength{\unitlength}{0.240900pt}
\ifx\plotpoint\undefined\newsavebox{\plotpoint}\fi
\sbox{\plotpoint}{\rule[-0.200pt]{0.400pt}{0.400pt}}%
\begin{picture}(1500,900)(0,0)
\font\gnuplot=cmr10 at 10pt
\gnuplot
\sbox{\plotpoint}{\rule[-0.200pt]{0.400pt}{0.400pt}}%
\put(176.0,68.0){\rule[-0.200pt]{303.534pt}{0.400pt}}
\put(176.0,68.0){\rule[-0.200pt]{0.400pt}{194.888pt}}
\put(176.0,68.0){\rule[-0.200pt]{4.818pt}{0.400pt}}
\put(154,68){\makebox(0,0)[r]{$0$}}
\put(1416.0,68.0){\rule[-0.200pt]{4.818pt}{0.400pt}}
\put(176.0,203.0){\rule[-0.200pt]{4.818pt}{0.400pt}}
\put(154,203){\makebox(0,0)[r]{$0.1$}}
\put(1416.0,203.0){\rule[-0.200pt]{4.818pt}{0.400pt}}
\put(176.0,338.0){\rule[-0.200pt]{4.818pt}{0.400pt}}
\put(154,338){\makebox(0,0)[r]{$0.2$}}
\put(1416.0,338.0){\rule[-0.200pt]{4.818pt}{0.400pt}}
\put(176.0,473.0){\rule[-0.200pt]{4.818pt}{0.400pt}}
\put(154,473){\makebox(0,0)[r]{$0.3$}}
\put(1416.0,473.0){\rule[-0.200pt]{4.818pt}{0.400pt}}
\put(176.0,607.0){\rule[-0.200pt]{4.818pt}{0.400pt}}
\put(154,607){\makebox(0,0)[r]{$0.4$}}
\put(1416.0,607.0){\rule[-0.200pt]{4.818pt}{0.400pt}}
\put(176.0,742.0){\rule[-0.200pt]{4.818pt}{0.400pt}}
\put(154,742){\makebox(0,0)[r]{$0.5$}}
\put(1416.0,742.0){\rule[-0.200pt]{4.818pt}{0.400pt}}
\put(176.0,877.0){\rule[-0.200pt]{4.818pt}{0.400pt}}
\put(154,877){\makebox(0,0)[r]{$0.6$}}
\put(1416.0,877.0){\rule[-0.200pt]{4.818pt}{0.400pt}}
\put(176.0,68.0){\rule[-0.200pt]{0.400pt}{4.818pt}}
\put(176,23){\makebox(0,0){$0$}}
\put(176.0,857.0){\rule[-0.200pt]{0.400pt}{4.818pt}}
\put(334.0,68.0){\rule[-0.200pt]{0.400pt}{4.818pt}}
\put(334,23){\makebox(0,0){$5$}}
\put(334.0,857.0){\rule[-0.200pt]{0.400pt}{4.818pt}}
\put(491.0,68.0){\rule[-0.200pt]{0.400pt}{4.818pt}}
\put(491,23){\makebox(0,0){$10$}}
\put(491.0,857.0){\rule[-0.200pt]{0.400pt}{4.818pt}}
\put(649.0,68.0){\rule[-0.200pt]{0.400pt}{4.818pt}}
\put(649,23){\makebox(0,0){$15$}}
\put(649.0,857.0){\rule[-0.200pt]{0.400pt}{4.818pt}}
\put(806.0,68.0){\rule[-0.200pt]{0.400pt}{4.818pt}}
\put(806,23){\makebox(0,0){$20$}}
\put(806.0,857.0){\rule[-0.200pt]{0.400pt}{4.818pt}}
\put(964.0,68.0){\rule[-0.200pt]{0.400pt}{4.818pt}}
\put(964,23){\makebox(0,0){$25$}}
\put(964.0,857.0){\rule[-0.200pt]{0.400pt}{4.818pt}}
\put(1121.0,68.0){\rule[-0.200pt]{0.400pt}{4.818pt}}
\put(1121,23){\makebox(0,0){$30$}}
\put(1121.0,857.0){\rule[-0.200pt]{0.400pt}{4.818pt}}
\put(1279.0,68.0){\rule[-0.200pt]{0.400pt}{4.818pt}}
\put(1279,23){\makebox(0,0){$35$}}
\put(1279.0,857.0){\rule[-0.200pt]{0.400pt}{4.818pt}}
\put(1436.0,68.0){\rule[-0.200pt]{0.400pt}{4.818pt}}
\put(1436,23){\makebox(0,0){$40$}}
\put(1436.0,857.0){\rule[-0.200pt]{0.400pt}{4.818pt}}
\put(176.0,68.0){\rule[-0.200pt]{303.534pt}{0.400pt}}
\put(1436.0,68.0){\rule[-0.200pt]{0.400pt}{194.888pt}}
\put(176.0,877.0){\rule[-0.200pt]{303.534pt}{0.400pt}}
\put(176.0,68.0){\rule[-0.200pt]{0.400pt}{194.888pt}}
\put(1306,812){\makebox(0,0)[r]{1}}
\put(1328.0,812.0){\rule[-0.200pt]{15.899pt}{0.400pt}}
\put(208,423){\usebox{\plotpoint}}
\multiput(208.58,423.00)(0.494,1.105){29}{\rule{0.119pt}{0.975pt}}
\multiput(207.17,423.00)(16.000,32.976){2}{\rule{0.400pt}{0.488pt}}
\multiput(224.00,458.59)(1.601,0.477){7}{\rule{1.300pt}{0.115pt}}
\multiput(224.00,457.17)(12.302,5.000){2}{\rule{0.650pt}{0.400pt}}
\multiput(239.00,461.92)(0.732,-0.492){19}{\rule{0.682pt}{0.118pt}}
\multiput(239.00,462.17)(14.585,-11.000){2}{\rule{0.341pt}{0.400pt}}
\multiput(255.00,450.92)(0.497,-0.494){29}{\rule{0.500pt}{0.119pt}}
\multiput(255.00,451.17)(14.962,-16.000){2}{\rule{0.250pt}{0.400pt}}
\multiput(271.00,434.92)(0.497,-0.494){29}{\rule{0.500pt}{0.119pt}}
\multiput(271.00,435.17)(14.962,-16.000){2}{\rule{0.250pt}{0.400pt}}
\multiput(287.58,416.82)(0.494,-0.839){27}{\rule{0.119pt}{0.767pt}}
\multiput(286.17,418.41)(15.000,-23.409){2}{\rule{0.400pt}{0.383pt}}
\multiput(302.58,392.51)(0.494,-0.625){29}{\rule{0.119pt}{0.600pt}}
\multiput(301.17,393.75)(16.000,-18.755){2}{\rule{0.400pt}{0.300pt}}
\multiput(318.58,372.30)(0.494,-0.689){29}{\rule{0.119pt}{0.650pt}}
\multiput(317.17,373.65)(16.000,-20.651){2}{\rule{0.400pt}{0.325pt}}
\multiput(334.58,350.39)(0.497,-0.662){59}{\rule{0.120pt}{0.629pt}}
\multiput(333.17,351.69)(31.000,-39.694){2}{\rule{0.400pt}{0.315pt}}
\multiput(365.00,310.92)(0.499,-0.497){61}{\rule{0.500pt}{0.120pt}}
\multiput(365.00,311.17)(30.962,-32.000){2}{\rule{0.250pt}{0.400pt}}
\multiput(397.00,278.92)(0.534,-0.497){55}{\rule{0.528pt}{0.120pt}}
\multiput(397.00,279.17)(29.905,-29.000){2}{\rule{0.264pt}{0.400pt}}
\multiput(428.00,249.92)(0.640,-0.497){47}{\rule{0.612pt}{0.120pt}}
\multiput(428.00,250.17)(30.730,-25.000){2}{\rule{0.306pt}{0.400pt}}
\multiput(460.00,224.92)(0.778,-0.496){37}{\rule{0.720pt}{0.119pt}}
\multiput(460.00,225.17)(29.506,-20.000){2}{\rule{0.360pt}{0.400pt}}
\multiput(491.00,204.92)(0.895,-0.495){33}{\rule{0.811pt}{0.119pt}}
\multiput(491.00,205.17)(30.316,-18.000){2}{\rule{0.406pt}{0.400pt}}
\multiput(523.00,186.92)(1.044,-0.494){27}{\rule{0.927pt}{0.119pt}}
\multiput(523.00,187.17)(29.077,-15.000){2}{\rule{0.463pt}{0.400pt}}
\multiput(554.00,171.92)(1.250,-0.493){23}{\rule{1.085pt}{0.119pt}}
\multiput(554.00,172.17)(29.749,-13.000){2}{\rule{0.542pt}{0.400pt}}
\multiput(586.00,158.92)(1.439,-0.492){19}{\rule{1.227pt}{0.118pt}}
\multiput(586.00,159.17)(28.453,-11.000){2}{\rule{0.614pt}{0.400pt}}
\multiput(617.00,147.92)(1.642,-0.491){17}{\rule{1.380pt}{0.118pt}}
\multiput(617.00,148.17)(29.136,-10.000){2}{\rule{0.690pt}{0.400pt}}
\multiput(649.00,137.93)(2.013,-0.488){13}{\rule{1.650pt}{0.117pt}}
\multiput(649.00,138.17)(27.575,-8.000){2}{\rule{0.825pt}{0.400pt}}
\multiput(680.00,129.93)(2.399,-0.485){11}{\rule{1.929pt}{0.117pt}}
\multiput(680.00,130.17)(27.997,-7.000){2}{\rule{0.964pt}{0.400pt}}
\multiput(712.00,122.93)(2.751,-0.482){9}{\rule{2.167pt}{0.116pt}}
\multiput(712.00,123.17)(26.503,-6.000){2}{\rule{1.083pt}{0.400pt}}
\multiput(743.00,116.93)(2.841,-0.482){9}{\rule{2.233pt}{0.116pt}}
\multiput(743.00,117.17)(27.365,-6.000){2}{\rule{1.117pt}{0.400pt}}
\multiput(775.00,110.94)(4.429,-0.468){5}{\rule{3.200pt}{0.113pt}}
\multiput(775.00,111.17)(24.358,-4.000){2}{\rule{1.600pt}{0.400pt}}
\multiput(806.00,106.93)(3.493,-0.477){7}{\rule{2.660pt}{0.115pt}}
\multiput(806.00,107.17)(26.479,-5.000){2}{\rule{1.330pt}{0.400pt}}
\multiput(838.00,101.95)(6.714,-0.447){3}{\rule{4.233pt}{0.108pt}}
\multiput(838.00,102.17)(22.214,-3.000){2}{\rule{2.117pt}{0.400pt}}
\multiput(869.00,98.95)(6.937,-0.447){3}{\rule{4.367pt}{0.108pt}}
\multiput(869.00,99.17)(22.937,-3.000){2}{\rule{2.183pt}{0.400pt}}
\multiput(901.00,95.95)(6.714,-0.447){3}{\rule{4.233pt}{0.108pt}}
\multiput(901.00,96.17)(22.214,-3.000){2}{\rule{2.117pt}{0.400pt}}
\multiput(932.00,92.95)(6.937,-0.447){3}{\rule{4.367pt}{0.108pt}}
\multiput(932.00,93.17)(22.937,-3.000){2}{\rule{2.183pt}{0.400pt}}
\put(964,89.17){\rule{6.300pt}{0.400pt}}
\multiput(964.00,90.17)(17.924,-2.000){2}{\rule{3.150pt}{0.400pt}}
\put(995,87.17){\rule{6.500pt}{0.400pt}}
\multiput(995.00,88.17)(18.509,-2.000){2}{\rule{3.250pt}{0.400pt}}
\put(1027,85.17){\rule{6.300pt}{0.400pt}}
\multiput(1027.00,86.17)(17.924,-2.000){2}{\rule{3.150pt}{0.400pt}}
\put(1058,83.67){\rule{7.709pt}{0.400pt}}
\multiput(1058.00,84.17)(16.000,-1.000){2}{\rule{3.854pt}{0.400pt}}
\put(1090,82.17){\rule{6.300pt}{0.400pt}}
\multiput(1090.00,83.17)(17.924,-2.000){2}{\rule{3.150pt}{0.400pt}}
\put(1121,80.67){\rule{7.709pt}{0.400pt}}
\multiput(1121.00,81.17)(16.000,-1.000){2}{\rule{3.854pt}{0.400pt}}
\put(1153,79.67){\rule{7.468pt}{0.400pt}}
\multiput(1153.00,80.17)(15.500,-1.000){2}{\rule{3.734pt}{0.400pt}}
\put(1184,78.67){\rule{7.709pt}{0.400pt}}
\multiput(1184.00,79.17)(16.000,-1.000){2}{\rule{3.854pt}{0.400pt}}
\put(1216,77.67){\rule{7.468pt}{0.400pt}}
\multiput(1216.00,78.17)(15.500,-1.000){2}{\rule{3.734pt}{0.400pt}}
\put(1247,76.67){\rule{7.709pt}{0.400pt}}
\multiput(1247.00,77.17)(16.000,-1.000){2}{\rule{3.854pt}{0.400pt}}
\put(1279,75.67){\rule{7.468pt}{0.400pt}}
\multiput(1279.00,76.17)(15.500,-1.000){2}{\rule{3.734pt}{0.400pt}}
\put(1342,74.67){\rule{7.468pt}{0.400pt}}
\multiput(1342.00,75.17)(15.500,-1.000){2}{\rule{3.734pt}{0.400pt}}
\put(1373,73.67){\rule{7.709pt}{0.400pt}}
\multiput(1373.00,74.17)(16.000,-1.000){2}{\rule{3.854pt}{0.400pt}}
\put(1350,812){\circle*{18}}
\put(208,423){\circle*{18}}
\put(224,458){\circle*{18}}
\put(239,463){\circle*{18}}
\put(255,452){\circle*{18}}
\put(271,436){\circle*{18}}
\put(287,420){\circle*{18}}
\put(302,395){\circle*{18}}
\put(318,375){\circle*{18}}
\put(334,353){\circle*{18}}
\put(365,312){\circle*{18}}
\put(397,280){\circle*{18}}
\put(428,251){\circle*{18}}
\put(460,226){\circle*{18}}
\put(491,206){\circle*{18}}
\put(523,188){\circle*{18}}
\put(554,173){\circle*{18}}
\put(586,160){\circle*{18}}
\put(617,149){\circle*{18}}
\put(649,139){\circle*{18}}
\put(680,131){\circle*{18}}
\put(712,124){\circle*{18}}
\put(743,118){\circle*{18}}
\put(775,112){\circle*{18}}
\put(806,108){\circle*{18}}
\put(838,103){\circle*{18}}
\put(869,100){\circle*{18}}
\put(901,97){\circle*{18}}
\put(932,94){\circle*{18}}
\put(964,91){\circle*{18}}
\put(995,89){\circle*{18}}
\put(1027,87){\circle*{18}}
\put(1058,85){\circle*{18}}
\put(1090,84){\circle*{18}}
\put(1121,82){\circle*{18}}
\put(1153,81){\circle*{18}}
\put(1184,80){\circle*{18}}
\put(1216,79){\circle*{18}}
\put(1247,78){\circle*{18}}
\put(1279,77){\circle*{18}}
\put(1310,76){\circle*{18}}
\put(1342,76){\circle*{18}}
\put(1373,75){\circle*{18}}
\put(1405,74){\circle*{18}}
\put(1310.0,76.0){\rule[-0.200pt]{7.709pt}{0.400pt}}
\sbox{\plotpoint}{\rule[-0.500pt]{1.000pt}{1.000pt}}%
\put(1306,767){\makebox(0,0)[r]{2}}
\multiput(1328,767)(20.756,0.000){4}{\usebox{\plotpoint}}
\put(1394,767){\usebox{\plotpoint}}
\put(208,389){\usebox{\plotpoint}}
\multiput(208,389)(5.896,19.900){3}{\usebox{\plotpoint}}
\multiput(224,443)(9.282,18.564){2}{\usebox{\plotpoint}}
\put(249.47,482.16){\usebox{\plotpoint}}
\put(267.29,492.38){\usebox{\plotpoint}}
\put(286.64,499.86){\usebox{\plotpoint}}
\multiput(287,500)(18.808,-8.777){0}{\usebox{\plotpoint}}
\put(305.75,492.30){\usebox{\plotpoint}}
\put(325.92,487.53){\usebox{\plotpoint}}
\multiput(334,485)(18.683,-9.040){2}{\usebox{\plotpoint}}
\put(383.13,463.20){\usebox{\plotpoint}}
\multiput(397,458)(18.444,-9.519){2}{\usebox{\plotpoint}}
\multiput(428,442)(19.015,-8.319){2}{\usebox{\plotpoint}}
\put(477.01,419.22){\usebox{\plotpoint}}
\multiput(491,412)(19.229,-7.812){2}{\usebox{\plotpoint}}
\multiput(523,399)(19.561,-6.941){2}{\usebox{\plotpoint}}
\put(573.29,380.77){\usebox{\plotpoint}}
\multiput(586,376)(19.141,-8.027){2}{\usebox{\plotpoint}}
\put(631.26,358.10){\usebox{\plotpoint}}
\multiput(649,352)(19.561,-6.941){2}{\usebox{\plotpoint}}
\multiput(680,341)(19.811,-6.191){2}{\usebox{\plotpoint}}
\put(730.01,326.35){\usebox{\plotpoint}}
\multiput(743,323)(20.136,-5.034){2}{\usebox{\plotpoint}}
\put(790.58,311.99){\usebox{\plotpoint}}
\multiput(806,309)(20.136,-5.034){2}{\usebox{\plotpoint}}
\put(851.32,298.42){\usebox{\plotpoint}}
\multiput(869,295)(20.507,-3.204){2}{\usebox{\plotpoint}}
\put(912.58,287.38){\usebox{\plotpoint}}
\multiput(932,283)(20.136,-5.034){2}{\usebox{\plotpoint}}
\multiput(964,275)(20.659,-1.999){2}{\usebox{\plotpoint}}
\put(1014.51,268.95){\usebox{\plotpoint}}
\multiput(1027,267)(20.491,-3.305){2}{\usebox{\plotpoint}}
\put(1075.91,258.64){\usebox{\plotpoint}}
\multiput(1090,256)(20.491,-3.305){2}{\usebox{\plotpoint}}
\put(1137.33,248.45){\usebox{\plotpoint}}
\multiput(1153,246)(20.712,-1.336){2}{\usebox{\plotpoint}}
\put(1199.08,241.17){\usebox{\plotpoint}}
\multiput(1216,238)(20.745,-0.669){2}{\usebox{\plotpoint}}
\put(1260.87,234.83){\usebox{\plotpoint}}
\multiput(1279,232)(20.491,-3.305){2}{\usebox{\plotpoint}}
\put(1322.46,225.83){\usebox{\plotpoint}}
\multiput(1342,224)(20.745,-0.669){2}{\usebox{\plotpoint}}
\multiput(1373,223)(20.507,-3.204){2}{\usebox{\plotpoint}}
\put(1405,218){\usebox{\plotpoint}}
\sbox{\plotpoint}{\rule[-0.200pt]{0.400pt}{0.400pt}}%
\put(1306,722){\makebox(0,0)[r]{3}}
\put(1328.0,722.0){\rule[-0.200pt]{15.899pt}{0.400pt}}
\put(224,338){\usebox{\plotpoint}}
\multiput(239.00,338.60)(2.236,0.468){5}{\rule{1.700pt}{0.113pt}}
\multiput(239.00,337.17)(12.472,4.000){2}{\rule{0.850pt}{0.400pt}}
\put(255,341.67){\rule{3.854pt}{0.400pt}}
\multiput(255.00,341.17)(8.000,1.000){2}{\rule{1.927pt}{0.400pt}}
\put(271,343.17){\rule{3.300pt}{0.400pt}}
\multiput(271.00,342.17)(9.151,2.000){2}{\rule{1.650pt}{0.400pt}}
\multiput(287.00,345.59)(1.304,0.482){9}{\rule{1.100pt}{0.116pt}}
\multiput(287.00,344.17)(12.717,6.000){2}{\rule{0.550pt}{0.400pt}}
\put(302,350.67){\rule{3.854pt}{0.400pt}}
\multiput(302.00,350.17)(8.000,1.000){2}{\rule{1.927pt}{0.400pt}}
\put(224.0,338.0){\rule[-0.200pt]{3.613pt}{0.400pt}}
\put(350,352.17){\rule{6.300pt}{0.400pt}}
\multiput(350.00,351.17)(17.924,2.000){2}{\rule{3.150pt}{0.400pt}}
\multiput(381.00,352.93)(2.399,-0.485){11}{\rule{1.929pt}{0.117pt}}
\multiput(381.00,353.17)(27.997,-7.000){2}{\rule{0.964pt}{0.400pt}}
\put(413,345.17){\rule{6.300pt}{0.400pt}}
\multiput(413.00,346.17)(17.924,-2.000){2}{\rule{3.150pt}{0.400pt}}
\multiput(444.00,343.93)(1.834,-0.489){15}{\rule{1.522pt}{0.118pt}}
\multiput(444.00,344.17)(28.841,-9.000){2}{\rule{0.761pt}{0.400pt}}
\put(476,334.67){\rule{7.468pt}{0.400pt}}
\multiput(476.00,335.17)(15.500,-1.000){2}{\rule{3.734pt}{0.400pt}}
\multiput(507.00,333.93)(2.399,-0.485){11}{\rule{1.929pt}{0.117pt}}
\multiput(507.00,334.17)(27.997,-7.000){2}{\rule{0.964pt}{0.400pt}}
\multiput(539.00,326.93)(2.013,-0.488){13}{\rule{1.650pt}{0.117pt}}
\multiput(539.00,327.17)(27.575,-8.000){2}{\rule{0.825pt}{0.400pt}}
\multiput(570.00,318.93)(2.399,-0.485){11}{\rule{1.929pt}{0.117pt}}
\multiput(570.00,319.17)(27.997,-7.000){2}{\rule{0.964pt}{0.400pt}}
\multiput(602.00,311.93)(2.751,-0.482){9}{\rule{2.167pt}{0.116pt}}
\multiput(602.00,312.17)(26.503,-6.000){2}{\rule{1.083pt}{0.400pt}}
\multiput(633.00,305.93)(3.493,-0.477){7}{\rule{2.660pt}{0.115pt}}
\multiput(633.00,306.17)(26.479,-5.000){2}{\rule{1.330pt}{0.400pt}}
\multiput(665.00,300.93)(1.776,-0.489){15}{\rule{1.478pt}{0.118pt}}
\multiput(665.00,301.17)(27.933,-9.000){2}{\rule{0.739pt}{0.400pt}}
\multiput(696.00,291.94)(4.575,-0.468){5}{\rule{3.300pt}{0.113pt}}
\multiput(696.00,292.17)(25.151,-4.000){2}{\rule{1.650pt}{0.400pt}}
\multiput(728.00,287.95)(6.714,-0.447){3}{\rule{4.233pt}{0.108pt}}
\multiput(728.00,288.17)(22.214,-3.000){2}{\rule{2.117pt}{0.400pt}}
\multiput(759.00,284.93)(2.841,-0.482){9}{\rule{2.233pt}{0.116pt}}
\multiput(759.00,285.17)(27.365,-6.000){2}{\rule{1.117pt}{0.400pt}}
\multiput(791.00,278.93)(3.382,-0.477){7}{\rule{2.580pt}{0.115pt}}
\multiput(791.00,279.17)(25.645,-5.000){2}{\rule{1.290pt}{0.400pt}}
\multiput(822.00,273.93)(1.834,-0.489){15}{\rule{1.522pt}{0.118pt}}
\multiput(822.00,274.17)(28.841,-9.000){2}{\rule{0.761pt}{0.400pt}}
\multiput(854.00,264.94)(4.429,-0.468){5}{\rule{3.200pt}{0.113pt}}
\multiput(854.00,265.17)(24.358,-4.000){2}{\rule{1.600pt}{0.400pt}}
\multiput(885.00,260.94)(4.575,-0.468){5}{\rule{3.300pt}{0.113pt}}
\multiput(885.00,261.17)(25.151,-4.000){2}{\rule{1.650pt}{0.400pt}}
\put(917,256.67){\rule{7.468pt}{0.400pt}}
\multiput(917.00,257.17)(15.500,-1.000){2}{\rule{3.734pt}{0.400pt}}
\multiput(948.00,255.93)(2.399,-0.485){11}{\rule{1.929pt}{0.117pt}}
\multiput(948.00,256.17)(27.997,-7.000){2}{\rule{0.964pt}{0.400pt}}
\put(980,248.67){\rule{7.468pt}{0.400pt}}
\multiput(980.00,249.17)(15.500,-1.000){2}{\rule{3.734pt}{0.400pt}}
\multiput(1011.00,247.93)(2.841,-0.482){9}{\rule{2.233pt}{0.116pt}}
\multiput(1011.00,248.17)(27.365,-6.000){2}{\rule{1.117pt}{0.400pt}}
\multiput(1043.00,241.93)(2.323,-0.485){11}{\rule{1.871pt}{0.117pt}}
\multiput(1043.00,242.17)(27.116,-7.000){2}{\rule{0.936pt}{0.400pt}}
\put(1074,234.67){\rule{7.709pt}{0.400pt}}
\multiput(1074.00,235.17)(16.000,-1.000){2}{\rule{3.854pt}{0.400pt}}
\multiput(1106.00,233.94)(4.429,-0.468){5}{\rule{3.200pt}{0.113pt}}
\multiput(1106.00,234.17)(24.358,-4.000){2}{\rule{1.600pt}{0.400pt}}
\put(1137,229.67){\rule{7.709pt}{0.400pt}}
\multiput(1137.00,230.17)(16.000,-1.000){2}{\rule{3.854pt}{0.400pt}}
\multiput(1169.00,228.94)(4.429,-0.468){5}{\rule{3.200pt}{0.113pt}}
\multiput(1169.00,229.17)(24.358,-4.000){2}{\rule{1.600pt}{0.400pt}}
\multiput(1200.00,224.94)(4.575,-0.468){5}{\rule{3.300pt}{0.113pt}}
\multiput(1200.00,225.17)(25.151,-4.000){2}{\rule{1.650pt}{0.400pt}}
\put(1232,220.17){\rule{6.300pt}{0.400pt}}
\multiput(1232.00,221.17)(17.924,-2.000){2}{\rule{3.150pt}{0.400pt}}
\multiput(1263.00,218.95)(6.937,-0.447){3}{\rule{4.367pt}{0.108pt}}
\multiput(1263.00,219.17)(22.937,-3.000){2}{\rule{2.183pt}{0.400pt}}
\multiput(1295.00,215.93)(3.382,-0.477){7}{\rule{2.580pt}{0.115pt}}
\multiput(1295.00,216.17)(25.645,-5.000){2}{\rule{1.290pt}{0.400pt}}
\put(1326,210.67){\rule{7.709pt}{0.400pt}}
\multiput(1326.00,211.17)(16.000,-1.000){2}{\rule{3.854pt}{0.400pt}}
\multiput(1358.00,209.94)(4.429,-0.468){5}{\rule{3.200pt}{0.113pt}}
\multiput(1358.00,210.17)(24.358,-4.000){2}{\rule{1.600pt}{0.400pt}}
\multiput(1389.00,205.93)(3.493,-0.477){7}{\rule{2.660pt}{0.115pt}}
\multiput(1389.00,206.17)(26.479,-5.000){2}{\rule{1.330pt}{0.400pt}}
\put(318.0,352.0){\rule[-0.200pt]{7.709pt}{0.400pt}}
\end{picture}\\[0.3cm]
$${\bf Fig. 2}$$ \\[1.0cm]

$${\bf Q^2F(Q^2)~~ vs ~~ Q^2}$$

\setlength{\unitlength}{0.240900pt}
\ifx\plotpoint\undefined\newsavebox{\plotpoint}\fi
\begin{picture}(1500,900)(0,0)
\font\gnuplot=cmr10 at 10pt
\gnuplot
\sbox{\plotpoint}{\rule[-0.200pt]{0.400pt}{0.400pt}}%
\put(176.0,68.0){\rule[-0.200pt]{303.534pt}{0.400pt}}
\put(176.0,68.0){\rule[-0.200pt]{0.400pt}{194.888pt}}
\put(176.0,68.0){\rule[-0.200pt]{4.818pt}{0.400pt}}
\put(154,68){\makebox(0,0)[r]{$0$}}
\put(1416.0,68.0){\rule[-0.200pt]{4.818pt}{0.400pt}}
\put(176.0,203.0){\rule[-0.200pt]{4.818pt}{0.400pt}}
\put(154,203){\makebox(0,0)[r]{$0.1$}}
\put(1416.0,203.0){\rule[-0.200pt]{4.818pt}{0.400pt}}
\put(176.0,338.0){\rule[-0.200pt]{4.818pt}{0.400pt}}
\put(154,338){\makebox(0,0)[r]{$0.2$}}
\put(1416.0,338.0){\rule[-0.200pt]{4.818pt}{0.400pt}}
\put(176.0,473.0){\rule[-0.200pt]{4.818pt}{0.400pt}}
\put(154,473){\makebox(0,0)[r]{$0.3$}}
\put(1416.0,473.0){\rule[-0.200pt]{4.818pt}{0.400pt}}
\put(176.0,607.0){\rule[-0.200pt]{4.818pt}{0.400pt}}
\put(154,607){\makebox(0,0)[r]{$0.4$}}
\put(1416.0,607.0){\rule[-0.200pt]{4.818pt}{0.400pt}}
\put(176.0,742.0){\rule[-0.200pt]{4.818pt}{0.400pt}}
\put(154,742){\makebox(0,0)[r]{$0.5$}}
\put(1416.0,742.0){\rule[-0.200pt]{4.818pt}{0.400pt}}
\put(176.0,877.0){\rule[-0.200pt]{4.818pt}{0.400pt}}
\put(154,877){\makebox(0,0)[r]{$0.6$}}
\put(1416.0,877.0){\rule[-0.200pt]{4.818pt}{0.400pt}}
\put(176.0,68.0){\rule[-0.200pt]{0.400pt}{4.818pt}}
\put(176,23){\makebox(0,0){$0$}}
\put(176.0,857.0){\rule[-0.200pt]{0.400pt}{4.818pt}}
\put(334.0,68.0){\rule[-0.200pt]{0.400pt}{4.818pt}}
\put(334,23){\makebox(0,0){$5$}}
\put(334.0,857.0){\rule[-0.200pt]{0.400pt}{4.818pt}}
\put(491.0,68.0){\rule[-0.200pt]{0.400pt}{4.818pt}}
\put(491,23){\makebox(0,0){$10$}}
\put(491.0,857.0){\rule[-0.200pt]{0.400pt}{4.818pt}}
\put(649.0,68.0){\rule[-0.200pt]{0.400pt}{4.818pt}}
\put(649,23){\makebox(0,0){$15$}}
\put(649.0,857.0){\rule[-0.200pt]{0.400pt}{4.818pt}}
\put(806.0,68.0){\rule[-0.200pt]{0.400pt}{4.818pt}}
\put(806,23){\makebox(0,0){$20$}}
\put(806.0,857.0){\rule[-0.200pt]{0.400pt}{4.818pt}}
\put(964.0,68.0){\rule[-0.200pt]{0.400pt}{4.818pt}}
\put(964,23){\makebox(0,0){$25$}}
\put(964.0,857.0){\rule[-0.200pt]{0.400pt}{4.818pt}}
\put(1121.0,68.0){\rule[-0.200pt]{0.400pt}{4.818pt}}
\put(1121,23){\makebox(0,0){$30$}}
\put(1121.0,857.0){\rule[-0.200pt]{0.400pt}{4.818pt}}
\put(1279.0,68.0){\rule[-0.200pt]{0.400pt}{4.818pt}}
\put(1279,23){\makebox(0,0){$35$}}
\put(1279.0,857.0){\rule[-0.200pt]{0.400pt}{4.818pt}}
\put(1436.0,68.0){\rule[-0.200pt]{0.400pt}{4.818pt}}
\put(1436,23){\makebox(0,0){$40$}}
\put(1436.0,857.0){\rule[-0.200pt]{0.400pt}{4.818pt}}
\put(176.0,68.0){\rule[-0.200pt]{303.534pt}{0.400pt}}
\put(1436.0,68.0){\rule[-0.200pt]{0.400pt}{194.888pt}}
\put(176.0,877.0){\rule[-0.200pt]{303.534pt}{0.400pt}}
\put(176.0,68.0){\rule[-0.200pt]{0.400pt}{194.888pt}}
\put(1306,812){\makebox(0,0)[r]{1}}
\put(1328.0,812.0){\rule[-0.200pt]{15.899pt}{0.400pt}}
\put(208,433){\usebox{\plotpoint}}
\multiput(208.58,433.00)(0.494,2.139){27}{\rule{0.119pt}{1.780pt}}
\multiput(207.17,433.00)(15.000,59.306){2}{\rule{0.400pt}{0.890pt}}
\multiput(223.58,496.00)(0.494,0.817){29}{\rule{0.119pt}{0.750pt}}
\multiput(222.17,496.00)(16.000,24.443){2}{\rule{0.400pt}{0.375pt}}
\multiput(239.00,522.58)(0.616,0.493){23}{\rule{0.592pt}{0.119pt}}
\multiput(239.00,521.17)(14.771,13.000){2}{\rule{0.296pt}{0.400pt}}
\multiput(255.00,533.95)(3.365,-0.447){3}{\rule{2.233pt}{0.108pt}}
\multiput(255.00,534.17)(11.365,-3.000){2}{\rule{1.117pt}{0.400pt}}
\multiput(271.00,530.93)(0.844,-0.489){15}{\rule{0.767pt}{0.118pt}}
\multiput(271.00,531.17)(13.409,-9.000){2}{\rule{0.383pt}{0.400pt}}
\multiput(286.00,521.93)(1.395,-0.482){9}{\rule{1.167pt}{0.116pt}}
\multiput(286.00,522.17)(13.579,-6.000){2}{\rule{0.583pt}{0.400pt}}
\multiput(302.58,514.20)(0.494,-0.721){29}{\rule{0.119pt}{0.675pt}}
\multiput(301.17,515.60)(16.000,-21.599){2}{\rule{0.400pt}{0.338pt}}
\multiput(318.00,492.92)(0.497,-0.494){29}{\rule{0.500pt}{0.119pt}}
\multiput(318.00,493.17)(14.962,-16.000){2}{\rule{0.250pt}{0.400pt}}
\multiput(334.58,475.70)(0.494,-0.566){27}{\rule{0.119pt}{0.553pt}}
\multiput(333.17,476.85)(15.000,-15.852){2}{\rule{0.400pt}{0.277pt}}
\multiput(349.58,458.51)(0.497,-0.625){61}{\rule{0.120pt}{0.600pt}}
\multiput(348.17,459.75)(32.000,-38.755){2}{\rule{0.400pt}{0.300pt}}
\multiput(381.00,419.92)(0.499,-0.497){59}{\rule{0.500pt}{0.120pt}}
\multiput(381.00,420.17)(29.962,-31.000){2}{\rule{0.250pt}{0.400pt}}
\multiput(412.58,387.87)(0.497,-0.515){61}{\rule{0.120pt}{0.512pt}}
\multiput(411.17,388.94)(32.000,-31.936){2}{\rule{0.400pt}{0.256pt}}
\multiput(444.00,355.92)(0.534,-0.497){55}{\rule{0.528pt}{0.120pt}}
\multiput(444.00,356.17)(29.905,-29.000){2}{\rule{0.264pt}{0.400pt}}
\multiput(475.00,326.92)(0.615,-0.497){49}{\rule{0.592pt}{0.120pt}}
\multiput(475.00,327.17)(30.771,-26.000){2}{\rule{0.296pt}{0.400pt}}
\multiput(507.00,300.92)(0.675,-0.496){43}{\rule{0.639pt}{0.120pt}}
\multiput(507.00,301.17)(29.673,-23.000){2}{\rule{0.320pt}{0.400pt}}
\multiput(538.00,277.92)(0.765,-0.496){39}{\rule{0.710pt}{0.119pt}}
\multiput(538.00,278.17)(30.527,-21.000){2}{\rule{0.355pt}{0.400pt}}
\multiput(570.00,256.92)(0.866,-0.495){33}{\rule{0.789pt}{0.119pt}}
\multiput(570.00,257.17)(29.363,-18.000){2}{\rule{0.394pt}{0.400pt}}
\multiput(601.00,238.92)(0.949,-0.495){31}{\rule{0.853pt}{0.119pt}}
\multiput(601.00,239.17)(30.230,-17.000){2}{\rule{0.426pt}{0.400pt}}
\multiput(633.00,221.92)(0.977,-0.494){29}{\rule{0.875pt}{0.119pt}}
\multiput(633.00,222.17)(29.184,-16.000){2}{\rule{0.438pt}{0.400pt}}
\multiput(664.00,205.92)(1.250,-0.493){23}{\rule{1.085pt}{0.119pt}}
\multiput(664.00,206.17)(29.749,-13.000){2}{\rule{0.542pt}{0.400pt}}
\multiput(696.00,192.92)(1.210,-0.493){23}{\rule{1.054pt}{0.119pt}}
\multiput(696.00,193.17)(28.813,-13.000){2}{\rule{0.527pt}{0.400pt}}
\multiput(727.00,179.93)(1.834,-0.489){15}{\rule{1.522pt}{0.118pt}}
\multiput(727.00,180.17)(28.841,-9.000){2}{\rule{0.761pt}{0.400pt}}
\multiput(759.00,170.92)(1.590,-0.491){17}{\rule{1.340pt}{0.118pt}}
\multiput(759.00,171.17)(28.219,-10.000){2}{\rule{0.670pt}{0.400pt}}
\multiput(790.00,160.93)(2.079,-0.488){13}{\rule{1.700pt}{0.117pt}}
\multiput(790.00,161.17)(28.472,-8.000){2}{\rule{0.850pt}{0.400pt}}
\multiput(822.00,152.93)(2.013,-0.488){13}{\rule{1.650pt}{0.117pt}}
\multiput(822.00,153.17)(27.575,-8.000){2}{\rule{0.825pt}{0.400pt}}
\multiput(853.00,144.93)(2.399,-0.485){11}{\rule{1.929pt}{0.117pt}}
\multiput(853.00,145.17)(27.997,-7.000){2}{\rule{0.964pt}{0.400pt}}
\multiput(885.00,137.93)(3.382,-0.477){7}{\rule{2.580pt}{0.115pt}}
\multiput(885.00,138.17)(25.645,-5.000){2}{\rule{1.290pt}{0.400pt}}
\multiput(916.00,132.93)(2.841,-0.482){9}{\rule{2.233pt}{0.116pt}}
\multiput(916.00,133.17)(27.365,-6.000){2}{\rule{1.117pt}{0.400pt}}
\multiput(948.00,126.93)(2.751,-0.482){9}{\rule{2.167pt}{0.116pt}}
\multiput(948.00,127.17)(26.503,-6.000){2}{\rule{1.083pt}{0.400pt}}
\multiput(979.00,120.94)(4.575,-0.468){5}{\rule{3.300pt}{0.113pt}}
\multiput(979.00,121.17)(25.151,-4.000){2}{\rule{1.650pt}{0.400pt}}
\multiput(1011.00,116.94)(4.429,-0.468){5}{\rule{3.200pt}{0.113pt}}
\multiput(1011.00,117.17)(24.358,-4.000){2}{\rule{1.600pt}{0.400pt}}
\multiput(1042.00,112.94)(4.575,-0.468){5}{\rule{3.300pt}{0.113pt}}
\multiput(1042.00,113.17)(25.151,-4.000){2}{\rule{1.650pt}{0.400pt}}
\multiput(1074.00,108.95)(6.714,-0.447){3}{\rule{4.233pt}{0.108pt}}
\multiput(1074.00,109.17)(22.214,-3.000){2}{\rule{2.117pt}{0.400pt}}
\multiput(1105.00,105.95)(6.937,-0.447){3}{\rule{4.367pt}{0.108pt}}
\multiput(1105.00,106.17)(22.937,-3.000){2}{\rule{2.183pt}{0.400pt}}
\multiput(1137.00,102.92)(1.315,-0.492){21}{\rule{1.133pt}{0.119pt}}
\multiput(1137.00,103.17)(28.648,-12.000){2}{\rule{0.567pt}{0.400pt}}
\multiput(1168.00,92.59)(2.399,0.485){11}{\rule{1.929pt}{0.117pt}}
\multiput(1168.00,91.17)(27.997,7.000){2}{\rule{0.964pt}{0.400pt}}
\multiput(1200.00,97.95)(6.714,-0.447){3}{\rule{4.233pt}{0.108pt}}
\multiput(1200.00,98.17)(22.214,-3.000){2}{\rule{2.117pt}{0.400pt}}
\put(1231,94.17){\rule{6.500pt}{0.400pt}}
\multiput(1231.00,95.17)(18.509,-2.000){2}{\rule{3.250pt}{0.400pt}}
\put(1263,92.17){\rule{6.300pt}{0.400pt}}
\multiput(1263.00,93.17)(17.924,-2.000){2}{\rule{3.150pt}{0.400pt}}
\put(1294,90.17){\rule{6.500pt}{0.400pt}}
\multiput(1294.00,91.17)(18.509,-2.000){2}{\rule{3.250pt}{0.400pt}}
\put(1326,88.67){\rule{7.468pt}{0.400pt}}
\multiput(1326.00,89.17)(15.500,-1.000){2}{\rule{3.734pt}{0.400pt}}
\put(1357,87.17){\rule{6.500pt}{0.400pt}}
\multiput(1357.00,88.17)(18.509,-2.000){2}{\rule{3.250pt}{0.400pt}}
\put(1389,85.67){\rule{7.468pt}{0.400pt}}
\multiput(1389.00,86.17)(15.500,-1.000){2}{\rule{3.734pt}{0.400pt}}
\put(1350,812){\circle*{18}}
\put(208,433){\circle*{18}}
\put(223,496){\circle*{18}}
\put(239,522){\circle*{18}}
\put(255,535){\circle*{18}}
\put(271,532){\circle*{18}}
\put(286,523){\circle*{18}}
\put(302,517){\circle*{18}}
\put(318,494){\circle*{18}}
\put(334,478){\circle*{18}}
\put(349,461){\circle*{18}}
\put(381,421){\circle*{18}}
\put(412,390){\circle*{18}}
\put(444,357){\circle*{18}}
\put(475,328){\circle*{18}}
\put(507,302){\circle*{18}}
\put(538,279){\circle*{18}}
\put(570,258){\circle*{18}}
\put(601,240){\circle*{18}}
\put(633,223){\circle*{18}}
\put(664,207){\circle*{18}}
\put(696,194){\circle*{18}}
\put(727,181){\circle*{18}}
\put(759,172){\circle*{18}}
\put(790,162){\circle*{18}}
\put(822,154){\circle*{18}}
\put(853,146){\circle*{18}}
\put(885,139){\circle*{18}}
\put(916,134){\circle*{18}}
\put(948,128){\circle*{18}}
\put(979,122){\circle*{18}}
\put(1011,118){\circle*{18}}
\put(1042,114){\circle*{18}}
\put(1074,110){\circle*{18}}
\put(1105,107){\circle*{18}}
\put(1137,104){\circle*{18}}
\put(1168,92){\circle*{18}}
\put(1200,99){\circle*{18}}
\put(1231,96){\circle*{18}}
\put(1263,94){\circle*{18}}
\put(1294,92){\circle*{18}}
\put(1326,90){\circle*{18}}
\put(1357,89){\circle*{18}}
\put(1389,87){\circle*{18}}
\put(1420,86){\circle*{18}}
\sbox{\plotpoint}{\rule[-0.500pt]{1.000pt}{1.000pt}}%
\put(1306,767){\makebox(0,0)[r]{2}}
\multiput(1328,767)(20.756,0.000){4}{\usebox{\plotpoint}}
\put(1394,767){\usebox{\plotpoint}}
\put(208,381){\usebox{\plotpoint}}
\multiput(208,381)(4.019,20.363){4}{\usebox{\plotpoint}}
\multiput(223,457)(6.104,19.838){3}{\usebox{\plotpoint}}
\multiput(239,509)(7.546,19.335){2}{\usebox{\plotpoint}}
\put(260.57,559.39){\usebox{\plotpoint}}
\multiput(271,577)(12.064,16.889){2}{\usebox{\plotpoint}}
\put(295.69,610.72){\usebox{\plotpoint}}
\put(311.02,624.07){\usebox{\plotpoint}}
\put(330.36,631.09){\usebox{\plotpoint}}
\put(347.28,642.62){\usebox{\plotpoint}}
\put(367.55,644.00){\usebox{\plotpoint}}
\multiput(381,644)(20.712,1.336){2}{\usebox{\plotpoint}}
\put(429.68,644.34){\usebox{\plotpoint}}
\multiput(444,643)(20.246,-4.572){2}{\usebox{\plotpoint}}
\put(490.62,632.10){\usebox{\plotpoint}}
\multiput(507,628)(19.561,-6.941){2}{\usebox{\plotpoint}}
\multiput(538,617)(19.434,-7.288){2}{\usebox{\plotpoint}}
\put(589.34,600.63){\usebox{\plotpoint}}
\multiput(601,598)(19.811,-6.191){2}{\usebox{\plotpoint}}
\put(648.11,580.69){\usebox{\plotpoint}}
\multiput(664,573)(19.434,-7.288){2}{\usebox{\plotpoint}}
\multiput(696,561)(19.753,-6.372){2}{\usebox{\plotpoint}}
\put(746.04,547.43){\usebox{\plotpoint}}
\multiput(759,545)(19.561,-6.941){2}{\usebox{\plotpoint}}
\put(805.58,529.62){\usebox{\plotpoint}}
\multiput(822,525)(19.932,-5.787){2}{\usebox{\plotpoint}}
\multiput(853,516)(19.229,-7.812){2}{\usebox{\plotpoint}}
\put(904.81,501.08){\usebox{\plotpoint}}
\multiput(916,500)(19.229,-7.812){2}{\usebox{\plotpoint}}
\put(963.69,481.94){\usebox{\plotpoint}}
\multiput(979,477)(20.507,-3.204){2}{\usebox{\plotpoint}}
\put(1024.35,468.55){\usebox{\plotpoint}}
\multiput(1042,464)(18.330,-9.738){2}{\usebox{\plotpoint}}
\multiput(1074,447)(20.585,-2.656){2}{\usebox{\plotpoint}}
\put(1122.98,441.31){\usebox{\plotpoint}}
\multiput(1137,440)(20.377,-3.944){2}{\usebox{\plotpoint}}
\put(1184.12,429.97){\usebox{\plotpoint}}
\multiput(1200,426)(20.246,-4.572){2}{\usebox{\plotpoint}}
\put(1245.05,417.68){\usebox{\plotpoint}}
\multiput(1263,416)(19.561,-6.941){2}{\usebox{\plotpoint}}
\multiput(1294,405)(20.665,-1.937){2}{\usebox{\plotpoint}}
\put(1345.89,396.22){\usebox{\plotpoint}}
\multiput(1357,393)(20.745,-0.648){2}{\usebox{\plotpoint}}
\put(1407.09,387.33){\usebox{\plotpoint}}
\put(1420,384){\usebox{\plotpoint}}
\sbox{\plotpoint}{\rule[-0.200pt]{0.400pt}{0.400pt}}%
\put(1306,722){\makebox(0,0)[r]{3}}
\put(1328.0,722.0){\rule[-0.200pt]{15.899pt}{0.400pt}}
\put(208,491){\usebox{\plotpoint}}
\put(208,489.67){\rule{3.854pt}{0.400pt}}
\multiput(208.00,490.17)(8.000,-1.000){2}{\rule{1.927pt}{0.400pt}}
\multiput(224.00,488.92)(0.756,-0.491){17}{\rule{0.700pt}{0.118pt}}
\multiput(224.00,489.17)(13.547,-10.000){2}{\rule{0.350pt}{0.400pt}}
\multiput(239.00,478.92)(0.570,-0.494){25}{\rule{0.557pt}{0.119pt}}
\multiput(239.00,479.17)(14.844,-14.000){2}{\rule{0.279pt}{0.400pt}}
\multiput(255.00,464.92)(0.808,-0.491){17}{\rule{0.740pt}{0.118pt}}
\multiput(255.00,465.17)(14.464,-10.000){2}{\rule{0.370pt}{0.400pt}}
\multiput(271.00,454.92)(0.669,-0.492){21}{\rule{0.633pt}{0.119pt}}
\multiput(271.00,455.17)(14.685,-12.000){2}{\rule{0.317pt}{0.400pt}}
\multiput(287.00,442.93)(1.103,-0.485){11}{\rule{0.957pt}{0.117pt}}
\multiput(287.00,443.17)(13.013,-7.000){2}{\rule{0.479pt}{0.400pt}}
\multiput(302.00,435.93)(1.712,-0.477){7}{\rule{1.380pt}{0.115pt}}
\multiput(302.00,436.17)(13.136,-5.000){2}{\rule{0.690pt}{0.400pt}}
\multiput(318.00,430.95)(3.365,-0.447){3}{\rule{2.233pt}{0.108pt}}
\multiput(318.00,431.17)(11.365,-3.000){2}{\rule{1.117pt}{0.400pt}}
\multiput(334.00,427.93)(1.712,-0.477){7}{\rule{1.380pt}{0.115pt}}
\multiput(334.00,428.17)(13.136,-5.000){2}{\rule{0.690pt}{0.400pt}}
\put(350,422.17){\rule{6.300pt}{0.400pt}}
\multiput(350.00,423.17)(17.924,-2.000){2}{\rule{3.150pt}{0.400pt}}
\put(381,420.67){\rule{7.709pt}{0.400pt}}
\multiput(381.00,421.17)(16.000,-1.000){2}{\rule{3.854pt}{0.400pt}}
\put(444,420.67){\rule{7.709pt}{0.400pt}}
\multiput(444.00,420.17)(16.000,1.000){2}{\rule{3.854pt}{0.400pt}}
\multiput(476.00,422.61)(6.714,0.447){3}{\rule{4.233pt}{0.108pt}}
\multiput(476.00,421.17)(22.214,3.000){2}{\rule{2.117pt}{0.400pt}}
\multiput(507.00,423.95)(6.937,-0.447){3}{\rule{4.367pt}{0.108pt}}
\multiput(507.00,424.17)(22.937,-3.000){2}{\rule{2.183pt}{0.400pt}}
\multiput(539.00,420.94)(4.429,-0.468){5}{\rule{3.200pt}{0.113pt}}
\multiput(539.00,421.17)(24.358,-4.000){2}{\rule{1.600pt}{0.400pt}}
\put(570,416.67){\rule{7.709pt}{0.400pt}}
\multiput(570.00,417.17)(16.000,-1.000){2}{\rule{3.854pt}{0.400pt}}
\multiput(602.00,415.93)(2.751,-0.482){9}{\rule{2.167pt}{0.116pt}}
\multiput(602.00,416.17)(26.503,-6.000){2}{\rule{1.083pt}{0.400pt}}
\put(633,409.17){\rule{6.500pt}{0.400pt}}
\multiput(633.00,410.17)(18.509,-2.000){2}{\rule{3.250pt}{0.400pt}}
\put(665,407.67){\rule{7.468pt}{0.400pt}}
\multiput(665.00,408.17)(15.500,-1.000){2}{\rule{3.734pt}{0.400pt}}
\multiput(696.00,406.95)(6.937,-0.447){3}{\rule{4.367pt}{0.108pt}}
\multiput(696.00,407.17)(22.937,-3.000){2}{\rule{2.183pt}{0.400pt}}
\multiput(728.00,403.94)(4.429,-0.468){5}{\rule{3.200pt}{0.113pt}}
\multiput(728.00,404.17)(24.358,-4.000){2}{\rule{1.600pt}{0.400pt}}
\multiput(759.00,399.95)(6.937,-0.447){3}{\rule{4.367pt}{0.108pt}}
\multiput(759.00,400.17)(22.937,-3.000){2}{\rule{2.183pt}{0.400pt}}
\multiput(791.00,396.93)(2.323,-0.485){11}{\rule{1.871pt}{0.117pt}}
\multiput(791.00,397.17)(27.116,-7.000){2}{\rule{0.936pt}{0.400pt}}
\multiput(822.00,389.93)(3.493,-0.477){7}{\rule{2.660pt}{0.115pt}}
\multiput(822.00,390.17)(26.479,-5.000){2}{\rule{1.330pt}{0.400pt}}
\multiput(854.00,384.95)(6.714,-0.447){3}{\rule{4.233pt}{0.108pt}}
\multiput(854.00,385.17)(22.214,-3.000){2}{\rule{2.117pt}{0.400pt}}
\multiput(885.00,381.93)(3.493,-0.477){7}{\rule{2.660pt}{0.115pt}}
\multiput(885.00,382.17)(26.479,-5.000){2}{\rule{1.330pt}{0.400pt}}
\put(917,376.17){\rule{6.300pt}{0.400pt}}
\multiput(917.00,377.17)(17.924,-2.000){2}{\rule{3.150pt}{0.400pt}}
\multiput(948.00,374.93)(2.841,-0.482){9}{\rule{2.233pt}{0.116pt}}
\multiput(948.00,375.17)(27.365,-6.000){2}{\rule{1.117pt}{0.400pt}}
\multiput(980.00,368.92)(1.439,-0.492){19}{\rule{1.227pt}{0.118pt}}
\multiput(980.00,369.17)(28.453,-11.000){2}{\rule{0.614pt}{0.400pt}}
\multiput(1011.00,357.93)(2.841,-0.482){9}{\rule{2.233pt}{0.116pt}}
\multiput(1011.00,358.17)(27.365,-6.000){2}{\rule{1.117pt}{0.400pt}}
\multiput(1043.00,351.93)(2.323,-0.485){11}{\rule{1.871pt}{0.117pt}}
\multiput(1043.00,352.17)(27.116,-7.000){2}{\rule{0.936pt}{0.400pt}}
\multiput(1074.00,344.93)(2.841,-0.482){9}{\rule{2.233pt}{0.116pt}}
\multiput(1074.00,345.17)(27.365,-6.000){2}{\rule{1.117pt}{0.400pt}}
\multiput(1106.00,338.94)(2.090,-0.468){5}{\rule{1.600pt}{0.113pt}}
\multiput(1106.00,339.17)(11.679,-4.000){2}{\rule{0.800pt}{0.400pt}}
\multiput(1121.00,334.93)(3.493,-0.477){7}{\rule{2.660pt}{0.115pt}}
\multiput(1121.00,335.17)(26.479,-5.000){2}{\rule{1.330pt}{0.400pt}}
\multiput(1153.00,329.93)(2.323,-0.485){11}{\rule{1.871pt}{0.117pt}}
\multiput(1153.00,330.17)(27.116,-7.000){2}{\rule{0.936pt}{0.400pt}}
\multiput(1184.00,322.95)(6.937,-0.447){3}{\rule{4.367pt}{0.108pt}}
\multiput(1184.00,323.17)(22.937,-3.000){2}{\rule{2.183pt}{0.400pt}}
\multiput(1216.00,319.93)(3.382,-0.477){7}{\rule{2.580pt}{0.115pt}}
\multiput(1216.00,320.17)(25.645,-5.000){2}{\rule{1.290pt}{0.400pt}}
\multiput(1247.00,314.94)(4.575,-0.468){5}{\rule{3.300pt}{0.113pt}}
\multiput(1247.00,315.17)(25.151,-4.000){2}{\rule{1.650pt}{0.400pt}}
\put(1279,310.67){\rule{7.468pt}{0.400pt}}
\multiput(1279.00,311.17)(15.500,-1.000){2}{\rule{3.734pt}{0.400pt}}
\put(1310,309.17){\rule{6.500pt}{0.400pt}}
\multiput(1310.00,310.17)(18.509,-2.000){2}{\rule{3.250pt}{0.400pt}}
\put(1342,307.17){\rule{6.300pt}{0.400pt}}
\multiput(1342.00,308.17)(17.924,-2.000){2}{\rule{3.150pt}{0.400pt}}
\put(1373,305.67){\rule{7.709pt}{0.400pt}}
\multiput(1373.00,306.17)(16.000,-1.000){2}{\rule{3.854pt}{0.400pt}}
\put(1405,304.17){\rule{6.300pt}{0.400pt}}
\multiput(1405.00,305.17)(17.924,-2.000){2}{\rule{3.150pt}{0.400pt}}
\put(413.0,421.0){\rule[-0.200pt]{7.468pt}{0.400pt}}
\end{picture}\\[0.3 cm]
$$ {\bf Fig. 3}$$

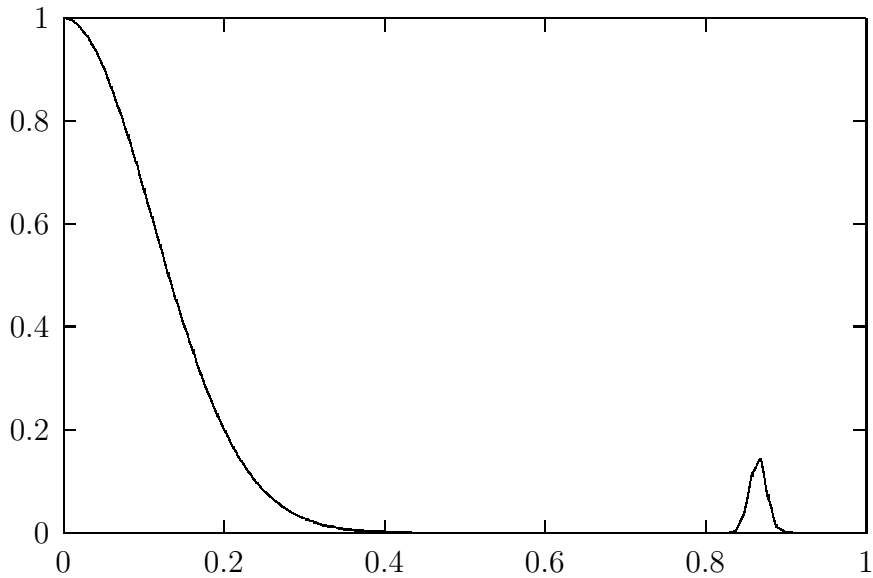
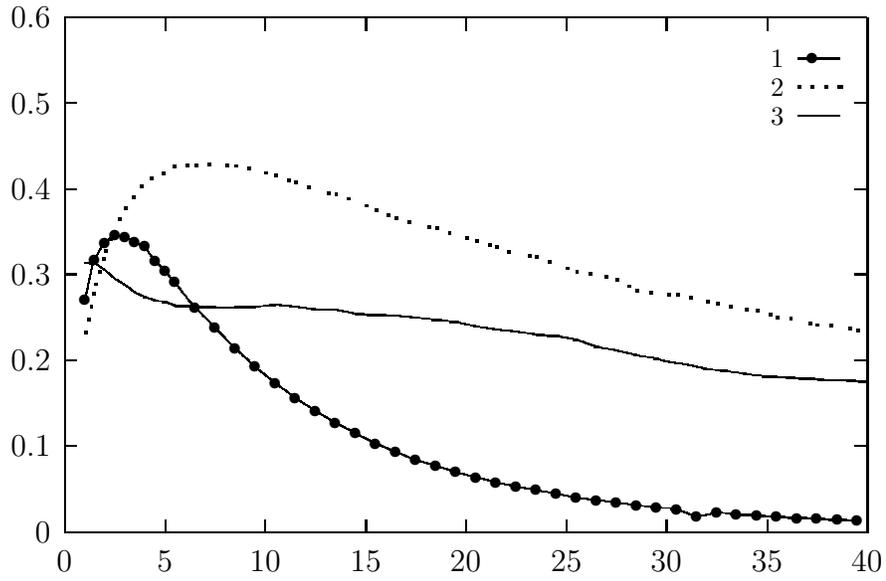
\end{document}